\documentclass[12pt]{article}
\usepackage{psfig}
\def\la{\mathrel{\mathpalette\fun <}}
\def\ga{\mathrel{\mathpalette\fun >}}
\def\fun#1#2{\lower3.6pt\vbox{\baselineskip0pt\lineskip.9pt
  \ialign{$\mathsurround=0pt#1\hfil##\hfil$\crcr#2\crcr\sim\crcr}}}

\begin{document}

\title{Ultrahigh Energy Neutrinos and Cosmic Rays as Probes of New Physics}

\author{G. Sigl\thanks{sigl@iap.fr}
\\[1cm]
{\normalsize
{\it Institut d'Astrophysique de Paris, C.N.R.S., 98 bis boulevard
Arago, F-75014 Paris, France}}}

\maketitle


\begin{abstract}
Cosmic high energy neutrinos are inextricably linked to the
origin of cosmic rays which is one of the major unresolved questions
in astrophysics. In particular,
the highest energy cosmic rays observed possess macroscopic energies and
their origin is likely to be associated with the most energetic processes
in the Universe. Their existence triggered a flurry of theoretical
explanations ranging from conventional shock acceleration to particle
physics beyond the Standard Model and processes taking place at the
earliest moments of our Universe. Furthermore, many new experimental
activities promise a considerable increase of statistics at the highest
energies and a combination with $\gamma-$ray and neutrino astrophysics
will put strong constraints on these theoretical models. The detection
of ultra high energy neutrinos in particular is made likely by
new experimental techniques and will open an important new channel.
We give an overview over this quickly evolving field with special
emphasize on new experimental ideas and possibilities for probing
new physics beyond the electroweak scale.
\end{abstract}

\vspace{6cm}




\newpage
\thispagestyle{empty}
\tableofcontents

\newpage
\setcounter{page}{1}


\section{Introduction}
The furtherst continuous source so far observed in neutrinos is the
Sun~\cite{solar}. These solar neutrinos together with
their atmospheric cousins~\cite{atmospheric,superk} which are produced
by cosmic ray interactions in the atmosphere have in fact provided
evidence for neutrino mass and oscillations. They are thus
playing an important role in probing new physics beyond the
Standard Model. The only extragalactic source so far observed
in neutrinos in the 10 MeV range was sporadic, namely supernova
1987A. These neutrinos have extensively been used to constrain
neutrino masses and oscillation parameters~\cite{sn1987a}.

Neutrinos at much higher energies, above a GeV or so,
have not been detected yet due to the small fluxes and cross
sections. However, the disadvantage of difficult detection
is at the same time a blessing because it makes these elusive
particles reach us unattenuated over cosmological distances and from
very dense environments where all other particles (except gravitational
waves) would be absorbed. Such neutrinos are expected
to be produced, apart from more speculative mechanisms such
as decay of superheavy particles, as secondary products of interactions
of the well known cosmic rays in extragalactic and galactic sources
as well as during propagation. These high energy neutrinos
are therefore inextricably linked with the physics and astrophysics
of cosmic rays. These lectures will therefore discuss high energy
neutrinos together with cosmic rays. The main focus will thereby
be on new possibilities opened by new experimental techniques
now under consideration or already under construction, and
resulting probes of new physics at very high energies beyond the
electroweak scale. We thus begin with a short discussion of the
now very active field of ultrahigh energy cosmic rays in general.

\begin{figure}[htb]
\centerline{\hbox{\psfig{figure=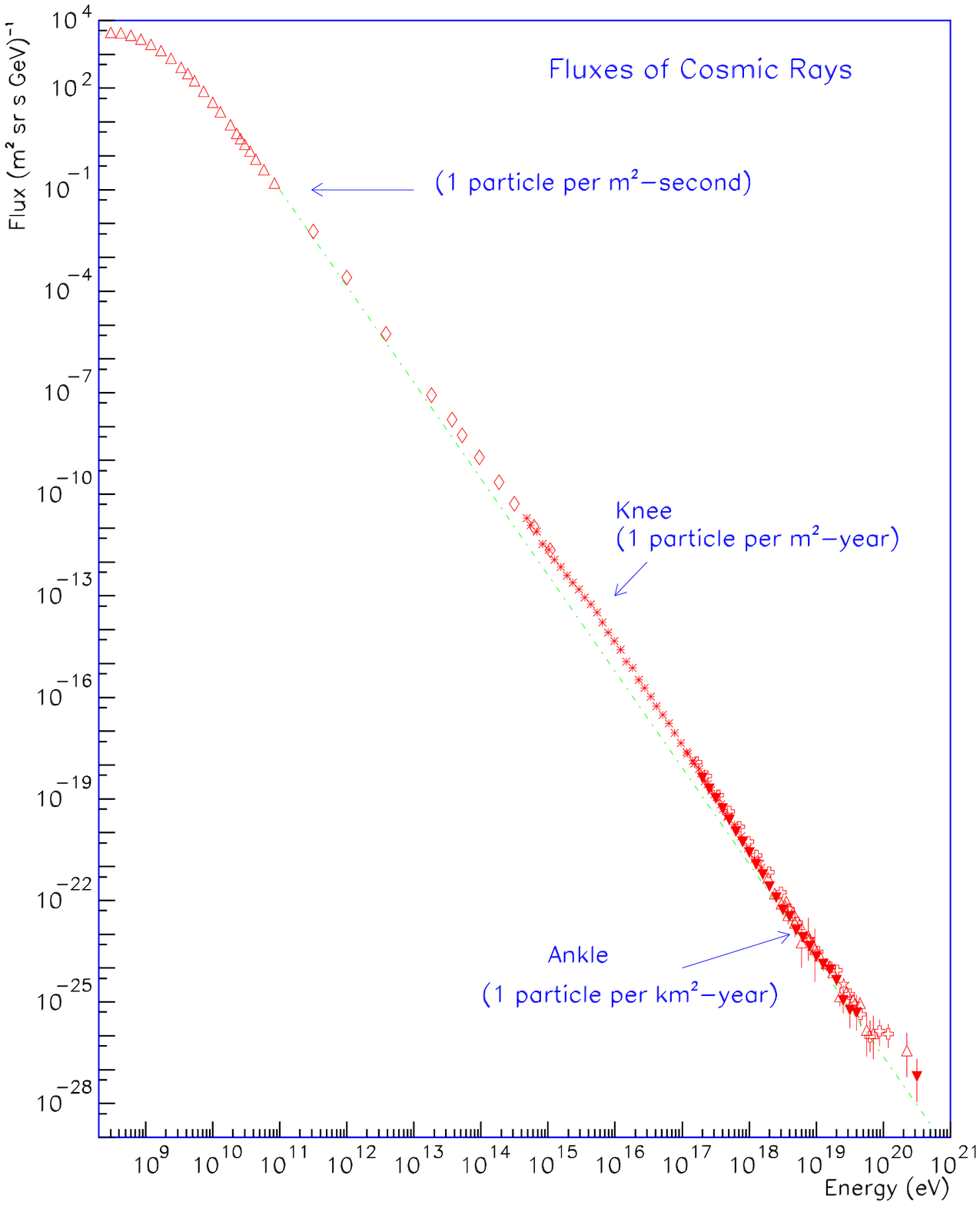,clip=true,height=0.9\textheight}}}
\caption[...]{The cosmic ray all particle
spectrum~\cite{swordy}. Approximate integral fluxes are also shown}
\label{fig1a}
\end{figure}

After almost 90 years of research on cosmic rays (CRs), their origin
is still an open question, for which the degree of uncertainty
increases with energy: Only below 100 MeV
kinetic energy, where the solar wind shields protons coming
from outside the solar system, the sun must give rise to
the observed proton flux. The bulk of the CRs up to at least an energy
of $E=4\times 10^{15}\,$eV is believed to originate within our Galaxy.
Above that energy, which is associated with the so called ``knee'',
the flux of particles per area, time, solid angle, and energy,
which can be well approximated by broken power laws
$\propto E^{-\gamma}$, steepens from a power law index $\gamma\simeq2.7$
to one of index $\simeq3.2$. Above the so called ``ankle'' at
$E\simeq5\times10^{18}\,$eV, the spectrum flattens again to a power law
of index $\gamma\simeq2.8$. This latter feature is often interpreted as a
cross over from a steeper Galactic component to a harder component
of extragalactic origin. Fig.~\ref{fig1a} shows the measured
CR spectrum above 100 MeV, up to $3\times10^{20}\,$eV, the highest
energy measured so far for an individual CR.

The conventional scenario
assumes that all high energy charged particles are accelerated in
magnetized astrophysical shocks, whose size and typical magnetic
field strength determines the maximal achievable energy, similar
to the situation in man made particle accelerators. The most
likely astrophysical accelerators for CR up to the knee, and
possibly up to the ankle are the shocks associated with
remnants of past Galactic supernova explosions, whereas for the
presumed extragalactic component powerful objects such as
active galactic nuclei are envisaged.

\begin{figure}[htb]
\centerline{\hbox{\psfig{figure=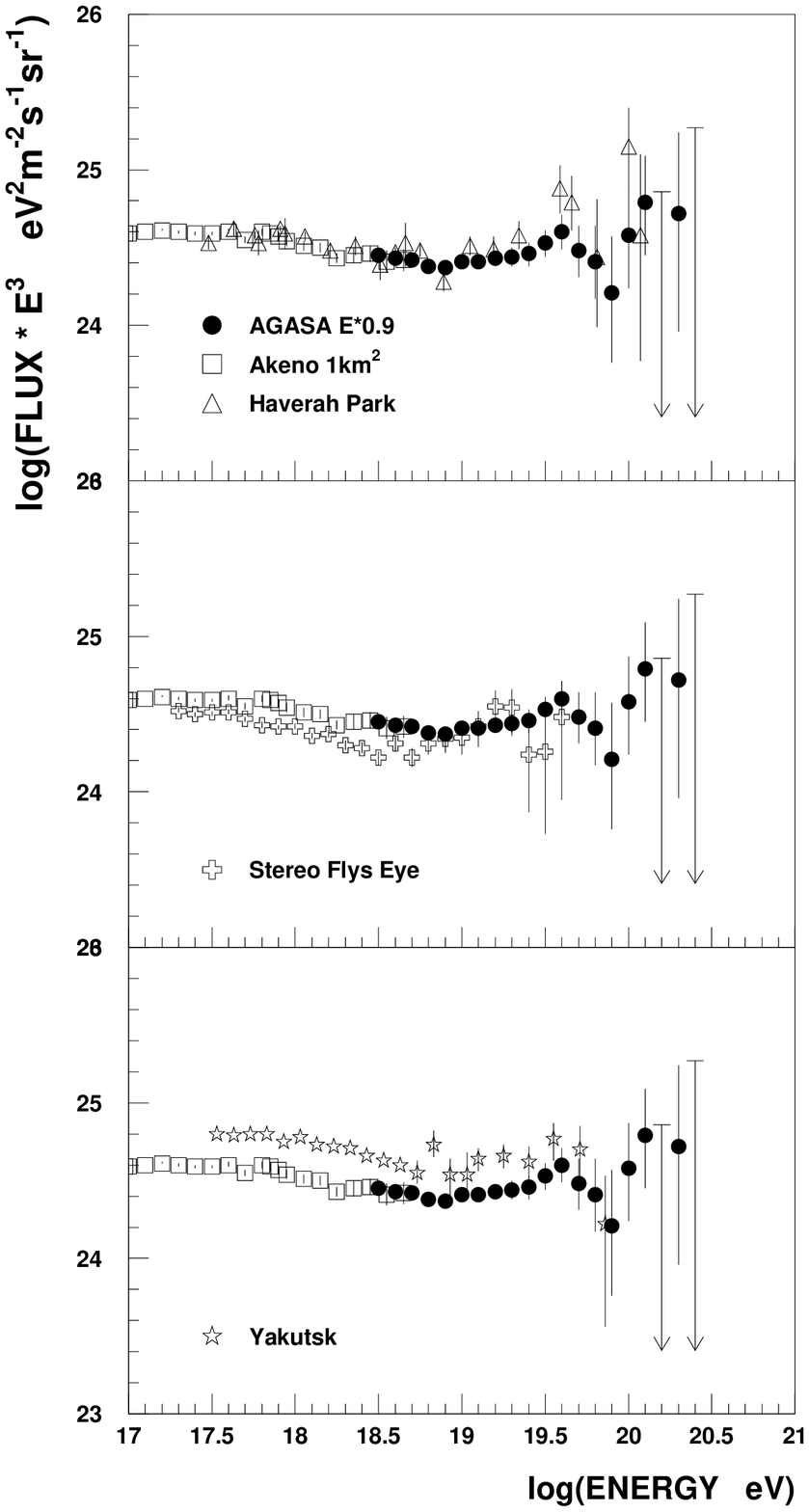,clip=true,height=0.9\textheight}}}
\caption[...]{The cosmic ray spectrum above $10^{17}\,$eV
(from Ref~\cite{review1}).The ``ankle'' is visible at
$E\simeq5\times10^{18}\,$eV.}
\label{fig1b}
\end{figure}

The main focus of this contribution will be on ultrahigh energy
cosmic rays (UHECRs), those with energy
$\ga10^{18}\,$eV~\cite{volcano,sugar,haverah,yakutsk,fe,agasa},
see Fig.~\ref{fig1b}, and neutrinos in the same energy range,
which have not been detected yet. For more details on CRs in general
the reader is referred to recent monographs~\cite{bbdgp,gaisser}.
In particular, extremely high energy (EHE)\footnote{We
shall use the abbreviation EHE to specifically denote energies
$E\ga10^{20}\,$eV, while the abbreviation UHE for ``Ultra-High Energy''
will sometimes be used to denote $E\ga$ 1 EeV, where 1 EeV =
$10^{18}\,$eV. Clearly UHE includes EHE but not vice versa.}
cosmic rays pose a serious challenge for conventional theories of
CR origin based on acceleration of charged particles in powerful
astrophysical objects. The question of the origin of these EHECRs
is, therefore, currently a subject of much intense debate and
discussions as well as experimental efforts; for a short review
for the non-experts see Ref.~\cite{science}, for recent brief
reviews see Ref.~\cite{review1,reviews} and for a detailed review
see Ref.~\cite{bs-rev}. In Sect.~2 we will summarize detection
techniques and present and future experimental projects.

The current theories of origin of EHECRs can be broadly categorized
into two distinct ``scenarios'': the ``bottom-up'' acceleration
scenario, and the ``top-down'' decay scenario, with various different models
within each scenario. As the names suggest, the
two scenarios are in a sense exact opposite of each other. The
bottom-up scenario is just an extension of the conventional
shock acceleration scenario in which charged particles are
accelerated from lower energies to the requisite high energies
in certain special astrophysical environments.
On the other hand, in the top-down scenario, the energetic
particles arise simply from decay of certain sufficiently massive
particles originating from physical processes in the early
Universe, and no acceleration mechanism is needed.

The problems encountered in trying to explain EHECRs in terms of
acceleration mechanisms have been well-documented in a number of studies;
see, e.g., Refs.~\cite{hillas-araa,ssb,norman}. Even if it is
possible, in principle, to accelerate particles to EHECR energies of order
100 EeV in some astrophysical sources, it is generally extremely difficult
to get the particles come out of the dense regions in and/or 
around the sources without losing much energy. Currently, the most
favorable sources in this
regard are perhaps a class of powerful radio galaxies (see, e.g., 
Refs.~\cite{bier-rev,kirk-duffy} for recent reviews and 
references to the literature), although the values of the relevant parameters 
required for acceleration to energies $\ga$ 100 EeV are somewhat on the
extreme side~\cite{norman}. However, even
if the requirements of energetics are met, the main problem with radio
galaxies as sources of EHECRs is that most of them seem to 
lie at large cosmological distances, $\gg$ 100 Mpc, from Earth. 
This is a major problem if EHECR
particles are conventional particles such as nucleons or heavy nuclei. The
reason is that nucleons above $\simeq$ 70 EeV lose energy drastically
during their propagation from the source to Earth due to the
Greisen-Zatsepin-Kuzmin (GZK) effect~\cite{greisen,zat-kuz}, namely, 
photo-production of pions when the nucleons collide with photons of the
cosmic microwave background (CMB), the mean-free path for which is $\sim$ 
few Mpc~\cite{stecker-gzk}. This process limits the possible
distance of any source of EHE nucleons to $\la$ 100 Mpc.
If the particles were heavy nuclei, 
they would be photo-disintegrated~\cite{psb,new_heavy} in
the CMB and infrared (IR) background within similar 
distances. Thus, nucleons or heavy
nuclei originating in distant radio galaxies are unlikely to survive with
EHECR energies at Earth with any significant flux, even if they were 
accelerated to energies of order 100 EeV at source. In addition,
if cosmic magnetic fields are not close to existing
upper limits, EHECRs are not likely to be deflected strongly
by large scale cosmological and/or Galactic magnetic fields
Thus, EHECR arrival directions should point back to their
sources in the sky and EHECRs
may offer us the unique opportunity of doing
charged particle astronomy. Yet, for the observed EHECR events so far, no
powerful sources close to the arrival directions of individual events
are found within about 100 Mpc~\cite{elb-som,ssb}. Very recently,
it has been suggested by Boldt and Ghosh~\cite{boldt-ghosh} that particles 
may be accelerated to energies $\sim10^{21}\,$eV near the event horizons of
spinning supermassive black holes associated with presently {\it inactive}
quasar remnants whose numbers within the local cosmological Universe
(i.e., within a GZK distance of order 50 Mpc) may be sufficient to explain
the observed EHECR flux. This would solve the problem of absence of
suitable currently {\it active} sources associated with EHECRs. A detailed
model incorporating this suggestion, however, remains to be worked out.

There are, of course, ways to avoid the distance restriction imposed
by the GZK effect, provided the problem of energetics is somehow
solved separately and provided one allows new physics beyond the
Standard Model of particle physics; we shall discuss suggestions
based on neutrinos in Sect.~3.

On the other hand, in the top-down scenario, which will be discussed
in Sect.~4, the problem of energetics is
trivially solved from the beginning. Here, the EHECR particles owe their
origin to decay of some supermassive ``X'' particles of mass
$m_X\gg10^{20}\,$eV, so that their decay products, envisaged as
the EHECR particles, can have energies all the way up to $\sim m_X$. Thus,
no acceleration mechanism is needed. The sources of the massive X
particles could be topological defects such as cosmic strings or magnetic
monopoles that could be produced in the early Universe during 
symmetry-breaking phase transitions envisaged in Grand Unified
Theories (GUTs). In an inflationary early Universe, the relevant
topological defects could be formed at a phase transition at the end of
inflation. Alternatively, the X particles could be certain
supermassive metastable relic particles of lifetime comparable to
or larger than the age of the Universe, which could be
produced in the early Universe through, for example, particle production
processes associated with inflation. Absence of nearby powerful
astrophysical objects such as AGNs or radio galaxies is not a problem in
the top-down scenario because the X particles or their sources need not
necessarily be associated with any specific active astrophysical objects.
In certain models, the X particles
themselves or their sources may be clustered in galactic halos, in
which case the dominant contribution to the EHECRs observed at Earth would
come from the X particles clustered within our Galactic Halo, for which 
the GZK restriction on source distance would be of no concern. 

By focusing primarily on ``non-conventional'' scenarios
involving new particle physics beyond the electroweak scale, we
do not wish to give the wrong impression that these scenarios explain all
aspects of EHECRs and UHE neutrinos. In fact, essentially each of the 
specific models that have been studied so
far has its own peculiar set of problems. Indeed, the main problem of
non-astrophysical solutions of the EHECR problem in general is that
they are highly model dependent.
On the other hand, it is precisely because of this
reason that these scenarios are also attractive --- they bring
in ideas of new physics beyond the Standard Model of particle physics
(such as Grand Unification and new interactions beyond the reach of
terrestrial accelerators) as well as ideas of early Universe cosmology
(such as topological defects and/or massive particle production in
inflation) into the realms of EHECRs where these ideas have the
potential to be tested by future EHECR experiments. 

The physics and astrophysics of UHECRs is not only intimately linked
with the emerging field of neutrino astronomy (for reviews see
Refs.~\cite{ghs,halzen}), but also with the already
established field of $\gamma-$ray astronomy (for reviews see, e.g.,
Ref.~\cite{gammarev}) which in turn are important
subdisciplines of particle astrophysics (for a review see, e.g.,
Ref.~\cite{mannheim}). Indeed, as we shall see, all
scenarios of UHECR origin, including the top-down models, are severely
constrained by neutrino and $\gamma-$ray observations and limits.
In turn, this linkage has important consequences for theoretical
predictions of fluxes of extragalactic neutrinos above a TeV
or so whose detection is a major goal of next-generation
neutrino telescopes (see Sect.~2): If these neutrinos are
produced as secondaries of protons accelerated in astrophysical
sources and if these protons are not absorbed in the sources,
but rather contribute to the UHECR flux observed, then
the energy content in the neutrino flux can not be higher
than the one in UHECRs, leading to the so called Waxman-Bahcall
bound~\cite{wb-bound,mpr}. If one of these assumptions
does not apply, such as for acceleration sources that are opaque
to nucleons or in the TD scenarios where X particle decays
produce much fewer nucleons than $\gamma-$rays and neutrinos,
the Waxman-Bahcall bound does not apply, but the neutrino
flux is still constrained by the observed diffuse $\gamma-$ray
flux in the GeV range.
This is true as long as the energy fluences produced in $\gamma-$rays
and neutrinos are comparable, which follows from isospin symmetry
if neutrinos are produced by pion production, because $\gamma-$rays
injected above the pair production threshold on the CMB will
cascade down to the GeV regime (see Sect.~4.4).

\section{Present and Future UHE CR and Neutrino Experiments}
The CR primaries are shielded by the Earth's
atmosphere and near the ground reveal their existence only by
indirect effects such as ionization. Indeed, it was the height
dependence of this latter effect which lead to the discovery of
CRs by Hess in 1912. Direct observation of CR primaries is only
possible from space by flying detectors with balloons or
spacecraft. Naturally, such detectors are very limited in size
and because the differential CR spectrum is a steeply falling function of 
energy (see Fig.~\ref{fig1a}), direct observations run out of
statistics typically around a few $100\,$TeV.

Above $\sim100\,$TeV, the showers of secondary particles
created in the interactions of the primary CR
with the atmosphere are extensive enough to be detectable from
the ground. In the most traditional technique, charged hadronic
particles, as well as electrons and muons in these Extensive Air
Showers (EAS) are recorded on the
ground~\cite{petrera} with standard instruments 
such as water Cherenkov detectors used in the old
Volcano Ranch~\cite{volcano} and Haverah Park~\cite{haverah}
experiments, and scintillation detectors which are
used now-a-days. Currently operating ground
arrays for UHECR EAS are the Yakutsk experiment
in Russia~\cite{yakutsk} and the Akeno
Giant Air Shower Array (AGASA) near Tokyo, Japan, which is
the largest one, covering an
area of roughly $100\,{\rm km}^2$ with about 100
detectors mutually separated by about $1\,$km~\cite{agasa}.
The Sydney University Giant Air Shower Recorder (SUGAR)~\cite{sugar}
operated until 1979 and was the largest array in the Southern
hemisphere. The ground array technique allows one to
measure a lateral cross section of the shower profile.
The energy of the shower-initiating primary particle is estimated by
appropriately parametrizing it in terms of a measurable parameter;
traditionally this parameter is taken to be the particle density at 600 m
from the shower core, which
is found to be quite insensitive to the primary composition
and the interaction model used to simulate air
showers.

The detection of secondary photons from EAS represents a
complementary technique. The experimentally most important light
sources are the fluorescence of air nitrogen excited by the charged
particles in the EAS and the Cherenkov radiation from the charged
particles that travel faster than the speed of light in the
atmospheric medium. The first
source is practically isotropic whereas the second one produces
light strongly concentrated on the surface of a cone around the
propagation direction of the charged source. The fluorescence
technique can be used equally well for both charged and neutral
primaries and was first used by the Fly's Eye detector~\cite{fe}
and will be part of several future projects on UHECRs
(see below). The primary energy can be estimated from
the total fluorescence yield. Information on the primary
composition is contained in the column depth $X_{\rm max}$ 
(measured in g$\,{\rm cm}^{-2}$) at which the shower reaches maximal
particle density. The average of $X_{\rm max}$ is related to the primary
energy $E$ by
\begin{equation}
  \left\langle X_{\rm max}\right\rangle = X_0^\prime\,\ln
  \left(\frac{E}{E_0}\right)\,.\label{elongation}
\end{equation}
Here, $X_0^\prime$ is called the elongation rate and $E_0$
is a characteristic energy that depends on the primary composition.
Therefore, if $X_{\rm max}$ and $X_0^\prime$ are  
determined from the longitudinal shower profile measured
by the fluorescence detector, then $E_0$ and thus
the composition, can be extracted after determining the energy
$E$ from the total fluorescence yield. 
Comparison of CR spectra measured with the ground array
and the fluorescence technique indicate systematic errors in
energy calibration that are generally smaller than $\sim$ 40\%.
For a more detailed discussion of experimental EAS analysis with the
ground array and the fluorescence technique see, e.g.,
Refs.~\cite{easrefs}.

As an upscaled version of the old Fly's Eye Cosmic Ray experiment, the
High Resolution Fly's Eye detector has started to take data
at Utah, USA~\cite{hires}. Taking into account a duty cycle of about
10\% (a fluorescence detector requires clear, moonless nights),
the effective aperture of this instrument will be
$\simeq350 (1000)\,{\rm km}^2\,{\rm sr}$ at $10 (100)\,$EeV,
on average about 6 times the Fly's Eye aperture, with a threshold
around $10^{17}\,$eV. Another project utilizing the fluorescence technique
is the Japanese Telescope Array~\cite{tel_array} which is currently
in the proposal stage. If approved, its effective aperture will
be about 10 times that of Fly's Eye above $10^{17}\,$eV, and
it would also be used as a Cherenkov detector for TeV $\gamma-$ray
astrophysics.
The largest project presently under construction is the  Pierre Auger
Giant Array Observatory~\cite{auger} planned for two sites, one in
Argentina and another in the USA for maximal sky coverage. Each site
will have  a $3000\,{\rm km}^2$ ground array. The southern
site will have about 1600 particle detectors (separated by 1.5 km
each) overlooked by four fluorescence
detectors. The ground arrays will have a duty cycle of nearly 100\%,
leading to an effective aperture about 30 times as large as the AGASA
array. The corresponding cosmic ray event rate above
$10^{20}\,$eV will be about 50 events per year. About 10\% 
of the events will be detected by both the ground array
and the fluorescence component and can be used for cross
calibration and detailed EAS studies. The energy threshold will
be around $10^{18}\,$eV, with full sensitivity above $10^{19}\,$eV.

Recently NASA initiated a concept study for detecting EAS
from space~\cite{owl,owl1} by observing their fluorescence light
from an Orbiting Wide-angle Light-collector (OWL). This would
provide an increase by another factor $\sim50$ in aperture
compared to the Pierre Auger Project, corresponding to an
event rate of up to a few thousand events per year above
$10^{20}\,$eV. Similar concepts such as the Extreme Universe
Space Observatory (EUSO)~\cite{euso} which is part
of the AirWatch program~\cite{airwatch} and of which a
prototype may be tested on the International Space Station
are also being discussed. It is possible that the OWL and AirWatch
efforts will merge. The energy threshold of such instruments
would be between $10^{19}$ and $10^{20}\,$eV. This technique
would be especially suitable for detection of very small
event rates such as those caused by UHE neutrinos which
would produce deeply penetrating EAS (see Sect.~3.2). For
more details on these recent experimental considerations
see Ref.~\cite{owl-proc}.

High energy neutrino astronomy is aiming towards a kilometer
scale neutrino observatory. The major technique is the optical
detection of Cherenkov light emitted by muons created in charged current
reactions of neutrinos with nucleons either in water
or in ice. The largest pilot experiments representing
these two detector media are the now defunct Deep Undersea Muon
and Neutrino Detection (DUMAND) experiment~\cite{dumand} in the
deep sea near Hawai and the Antarctic Muon And Neutrino Detector Array
(AMANDA) experiment~\cite{amanda} in the South
Pole ice. Another water based experiment is situated at
Lake Baikal~\cite{baikal}. Next generation deep sea projects
include the French Astronomy with a Neutrino Telescope and
Abyss environmental RESearch (ANTARES)~\cite{antares}
and the underwater Neutrino Experiment SouthwesT Of GReece
(NESTOR) project in the Mediterranean~\cite{nestor},
whereas ICECUBE~\cite{icecube} represents the planned kilometer scale
version of the AMANDA detector. Also under consideration are neutrino
detectors utilizing techniques to detect the radio pulse from the
electromagnetic showers created by neutrino
interactions in ice and other materials (see Ref.~\cite{radhep}).
This technique could possibly be scaled up to an effective area of
$10^4\,{\rm km}^2$ and a prototype is represented by the
Radio Ice Cherenkov Experiment (RICE) experiment at the
South Pole~\cite{rice}. The Goldstone radio telescope has
already put an upper limit on UHE neutrino fluxes~\cite{goldstone}
from the non-observation of radio pulses from showers induced by
neutrinos interacting in the moons rim~\cite{radio}.
The sensitivity of existing underwater acoustic arrays to
neutrino induced showers is also being studied~\cite{acoustic}.
Furthermore, neutrinos can also initiate horizontal EAS which can be
detected by giant ground arrays and fluorescence detectors
such as the Pierre Auger Project~\cite{auger-neut}. It has
also been shown that the sensitivity of such experiments to
UHE neutrinos could be significantly enhanced by triggering
on Earth-skimming showers, i.e. showers with zenith angles
slightly larger than $90^\circ$~\cite{fargion,auger-tau,upgoing},
especially for $\tau-$neutrinos.
Finally, as mentioned above, deeply penetrating EAS could be detected from
space by instruments such as the proposed space based
AirWatch type detectors~\cite{owl,owl1,euso,airwatch}.
More details and references on neutrino astronomy detectors are contained
in Refs.~\cite{ghs,learned}, and some recent overviews on
neutrino astronomy can be found in Ref.~\cite{halzen}.

\section{Neutrino Interactions and Propagation}

\subsection{Neutrino propagation}

The propagation of UHE neutrinos is governed mainly by their interaction
with the relic neutrino background (RNB). In this section we give a short
overview over the relevant interactions within the general framework
for particle propagation used in the present contribution.

The average squared CM energy for interaction of an UHE
neutrino of energy $E$ with a relic neutrino of energy
$\varepsilon$ is given by
\begin{equation}
  \left\langle s\right\rangle\simeq(45\,{\rm GeV})^2
  \left(\frac{\varepsilon}{10^{-3}\,{\rm eV}}\right)
  \left(\frac{E}{10^{15}\,{\rm GeV}}\right)\,.\label{snu}
\end{equation}
If the relic neutrino is relativistic, then
$\varepsilon\simeq3T_\nu(1+\eta_b/4)$ in Eq.~(\ref{snu}), where
$T_\nu\simeq1.9(1+z)\,{\rm K}=1.6\times10^{-4}(1+z)\,$eV is the
temperature at redshift $z$ and $\eta_b\la50$ is the dimensionless
chemical potential of relativistic relic neutrinos. For
nonrelativistic relic neutrinos of mass $m_\nu\la20\,$eV,
$\varepsilon\simeq\max\left[3T_\nu,m_\nu\right]$. Note that
Eq.~(\ref{snu}) implies interaction energies
that are typically smaller than electroweak energies even for UHE
neutrinos, except for energies near the Grand Unification scale,
$E\ga10^{15}\,$GeV, or if $m_\nu\ga1\,$eV. In this energy range,
the cross sections are given by the Standard Model of
electroweak interactions which are well confirmed
experimentally. Physics beyond the Standard Model is, therefore,
not expected to play a significant role in UHE neutrino interactions
with the low energy relic backgrounds.

The dominant interaction mode of UHE neutrinos with the RNB is the
exchange of a
$W^\pm$ boson in the t-channel ($\nu_i+\bar\nu_j\to l_i+\bar{l}_j$), 
or of a $Z^0$ boson in either the s-channel ($\nu_i+\bar\nu_i\to
f\bar{f}$) or the t-channel
($\nu_i+\bar\nu_j\to\nu_i+\bar\nu_j$)~\cite{weiler1,roulet,yoshida,ydjs}.
Here, $i,j$ stands for either the electron, muon, or tau flavor, where
$i\neq j$ for the first reaction, $l$ denotes a
charged lepton, and $f$ any charged fermion. If the latter is
a quark, it will, of course, subsequently fragment into hadrons.
As an example, the differential cross section for s-channel
production of $Z^0$ is given by
\begin{equation}
  \frac{d\sigma_{\nu_i+\bar\nu_j\to Z^0\to f\bar{f}}}{d\mu}
  =\frac{G_{\rm F}^2s}{4\pi}\,\frac{M_Z^2}{(s-M_Z^2)^2+M_Z^2
  \Gamma_Z^2}\left[g_L^2(1+\mu^*)^2+g_R^2(1-\mu^*)^2\right]
  \,,\label{Zres}
\end{equation}
where $G_{\rm F}$ is the Fermi constant, $M_Z$ and $\Gamma_Z$ are
mass and lifetime of the $Z^0$, $g_L$ and $g_R$ are the usual
dimensionless left- and right-handed coupling constants for $f$,
and $\mu^*$ is the cosine of the scattering angle in the CM system.

The t-channel processes have cross sections that rise linearly
with $s$ up to $s\simeq M_W^2$, with $M_W$ the $W^\pm$ mass,
above which they are roughly constant with a value
$\sigma_t(s\ga M_W)\sim G_{\rm F}^2M_W^2\sim10^{-34}\,{\rm cm}^2$.
Using Eq.~(\ref{snu}) this yields the rough estimate
\begin{eqnarray}
  \sigma_t(E,\varepsilon)&\sim&\min\left[10^{-34},10^{-44}
  \left(\frac{s}{{\rm MeV}^2}\right)\right]\,{\rm cm}^2
  \label{ewcross}\\
  &\sim&\min\left[10^{-34},3\times10^{-39}
  \left(\frac{\varepsilon}{10^{-3}\,{\rm eV}}\right)
  \left(\frac{E}{10^{20}\,{\rm eV}}\right)\right]\,{\rm cm}^2
  \,.\nonumber
\end{eqnarray}
In contrast, within the Standard Model the neutrino-nucleon cross
section roughly behaves as
\begin{equation}
  \sigma_{\nu N}(E)\sim10^{-31}(E/10^{20}
  \,{\rm eV})^{0.4}\,{\rm cm}^2\label{cccross1}
\end{equation}
for $E\ga10^{15}\,$eV (see Eq.(\ref{cccross2}) below).
Interactions of UHE neutrinos with nucleons are, however, still
negligible compared to interactions with the RNB because the
RNB particle density is about ten orders of magnitude larger 
than the baryon density. The only exception could occur near
Grand Unification scale energies and at high redshifts and/or
if contributions to the neutrino-nucleon cross section from
physics beyond the Standard Model dominate at these energies
(see Sect.~3.3 below).

\begin{figure}
\begin{center}
\centerline{\hbox{\psfig{figure=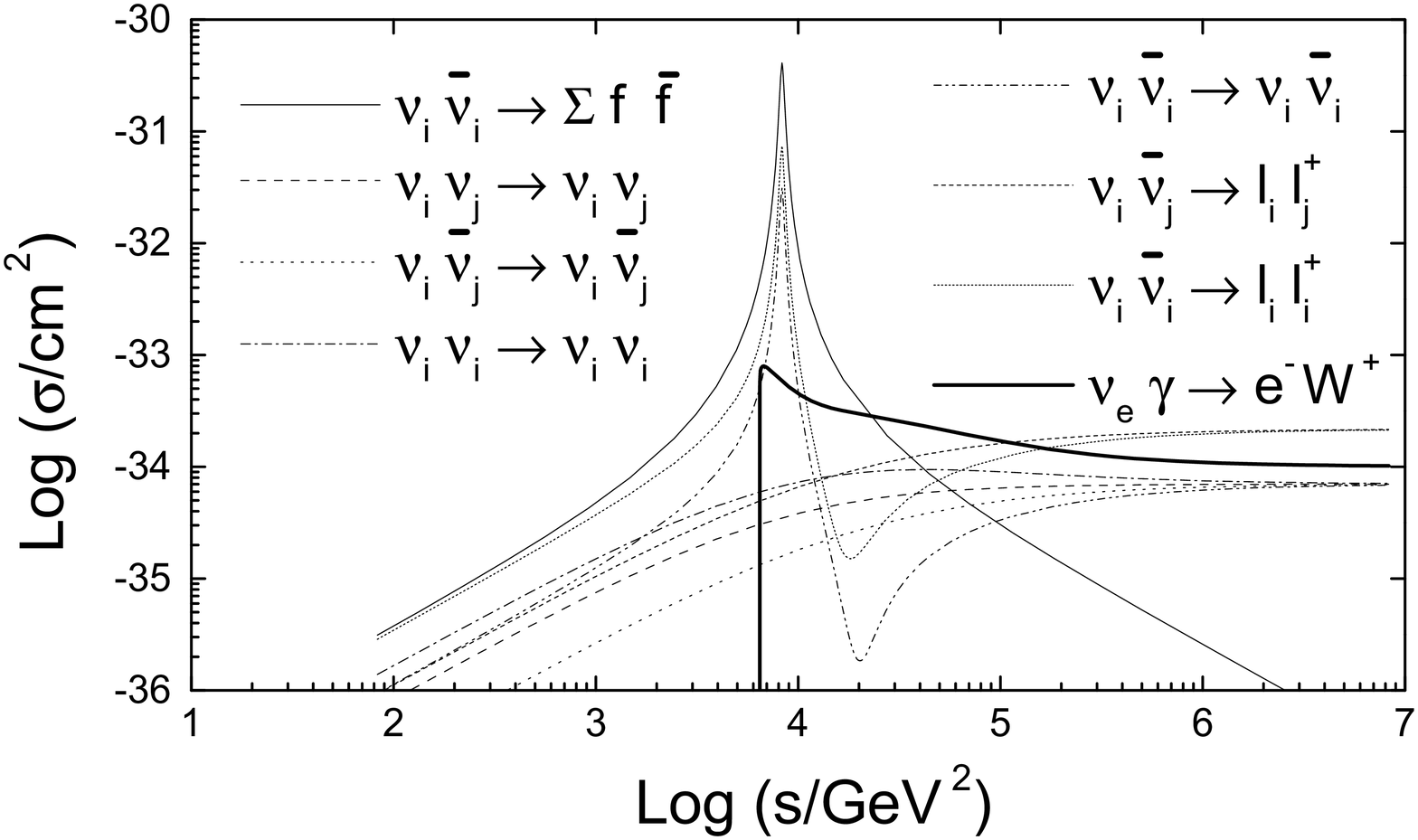,clip=true,width=0.9\textwidth}}}
\end{center}
\caption[...]{Various cross sections relevant for neutrino
propagation as a function of $s$~\cite{roulet,seckel}.  
The sum $\sum_jf_j\bar f_j$ does not include $f_j = \nu_i\,,l_i\,,t\,,W$,
or $Z$. (From Ref.~\cite{seckel}).}
\label{fig2}
\end{figure}

It has recently been pointed out~\cite{seckel} that above
the threshold for $W^\pm$ production the
process $\nu+\gamma\to lW^+$ becomes comparable to the
$\nu\nu$ processes discussed above. Fig.~\ref{fig2} compares
the cross sections relevant for neutrino propagation
at CM energies around the electroweak scale. Again,
for UHE neutrino interactions with the RNB
the relevant CM energies can only be reached if 
(a) the UHE neutrino energy is close to the Grand Unification scale,
or (b) the RNB neutrinos have masses in the eV regime, or (c) at
redshifts $z\ga10^3$. Even then the $\nu\gamma$ process
never dominates over the $\nu\nu$ process.

At lower energies
there is an additional $\nu\gamma$ interaction that was
recently discussed as potentially important besides the
$\nu\nu$ processes: Using an effective Lagrangian derived
from the Standard Model, Ref.~\cite{dr} obtained the
result $\sigma_{\gamma+\nu\to\gamma+\gamma+\nu}(s)\simeq
9\times10^{-56}\,(s/{\rm MeV}^2)^5\,{\rm cm}^2$, supposed
to be valid at least up to $s\la10\,{\rm MeV}^2$. Above
the electron pair production threshold the cross section
has not been calculated because of its complexity but is
likely to level off and eventually decrease. Nevertheless,
if the $s^5$ behavior holds up to $s\simeq$ a few hundred
MeV$^2$, comparison with Eq.~(\ref{ewcross}) shows that
the process $\gamma+\nu\to\gamma+\gamma+\nu$ would start
to dominate and influence neutrino propagation around
$E\sim3\times10^{17}\left(\varepsilon/10^{-3}\,{\rm eV}\right)
\,$eV, as was pointed out in Ref.~\cite{hwt}.

For a given source distribution, the contribution of the ``direct''
neutrinos to the flux can be computed from ($t_0$ is the age of the
Universe)
\begin{equation}
  j(E)\simeq\frac{3}{8\pi}t_0\int_0^{z(E)}dz_i(1+z_i)^{-9/2}
  \Phi[(1+z_i)E,z_i]\,,\label{cel_sol}
\end{equation}
up to the interaction redshift
$z(E)$, i.e. the average redshift from which a neutrino of
present day energy $E$ could have propagated without
interacting. This approximation neglects the secondary neutrinos
and the decay products of the leptons created
in the neutral current and charged current reactions of UHE
neutrinos with the RNB discussed above.
Similarly to the EM case, these secondary particles can lead to
neutrino cascades developing over cosmological
redshifts~\cite{yoshida}.

Approximate expressions for the interaction redshift for the
processes discussed above
have been given in Refs.~\cite{bbdgp,bhs} for CM energies
below the electroweak scale, assuming relativistic,
nondegenerate relic neutrinos, $m_\nu\la T_\nu$, and
$\eta_b\ll1$. Approaching the electroweak scale,
a resonance occurs in the interaction cross section for
s-channel $Z^0$ exchange at the $Z^0$ mass,
$s=M_Z^2\simeq(91\,{\rm GeV})^2$, see Eq.~(\ref{Zres}).
The absorption redshift for
the corresponding neutrino energy, $E\simeq10^{15}\,{\rm
GeV}(\varepsilon/10^{-3}\,{\rm eV})^{-1}$ drops to a few (or
less for a degenerate, relativistic RNB) and asymptotically
approaches constant values of a few tens at higher energies.

\begin{figure}
\begin{center}
\centerline{\hbox{\psfig{figure=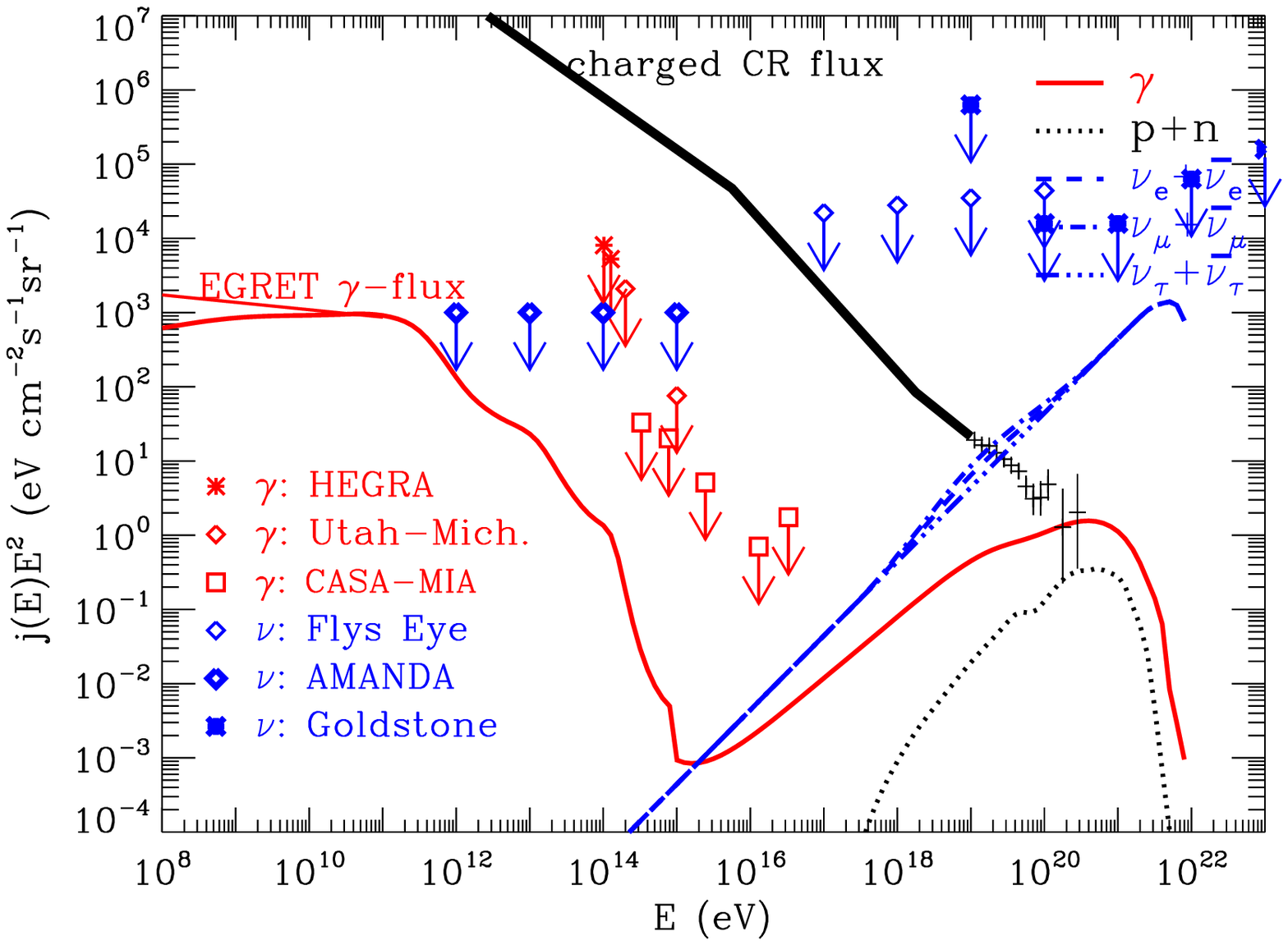,clip=true,width=0.9\textwidth}}}
\end{center}
\caption[...]{Fluxes of neutrinos (dashed and dashed-dotted, as
indicated), $\gamma-$rays (solid), and nucleons (dotted) predicted
by the Z-burst mechanism for $m_{\nu_e}=0.1\,$eV, $m_{\nu_\mu}=
m_{\nu_\tau}=1\,$eV, for homogeneously distributed sources
emitting neutrinos with an $E^{-1}$ spectrum (equal for all flavors) up to
$10^{22}\,$eV and an $E^{-2}$ $\gamma-$ray spectrum of equal
power up to 100 TeV. Injection rates were assumed comovingly constant
up to $z=2$. The relic neutrino overdensity was assumed to be
200 over 5 Mpc. The calculation
used the code described in Ref.~\cite{slby} and assumed the
lower limit of the universal radio background~\cite{pb} and
a vanishing extragalactic magnetic field.
1 sigma error bars are the combined data from the Haverah Park~\cite{haverah},
the Fly's Eye~\cite{fe}, and the AGASA~\cite{agasa} experiments
above $10^{19}\,$eV. Also shown are piecewise power law fits to the observed
charged CR flux (thick solid line) and the EGRET measurement
of the diffuse $\gamma-$ray flux between 30 MeV and 100 GeV~\cite{egret}
(solid line on left margin). Points with arrows represent upper
limits on the $\gamma-$ray flux from the HEGRA~\cite{hegra}, the
Utah-Michigan~\cite{utahmich}, and the CASA-MIA~\cite{casa}
experiments, and on the neutrino flux from the
Fly's Eye experiment~\cite{baltrusaitis}, the
Goldstone radio telescope~\cite{goldstone}, and the
AMANDA neutrino telescope~\cite{amanda}, as indicated
(see text and Ref.~\cite{bs-rev} for more details).}
\label{fig3}
\end{figure}

An interesting situation arises if the RNB consists of massive
neutrinos with $m_\nu\sim1\,$eV: Such neutrinos would constitute
hot dark matter which is expected to cluster~\cite{cowsik-mccl}, 
for example, in galaxy clusters. This would potentially increase 
the interaction probability for any neutrino of energy within the
width of the $Z^0$ resonance at $E=M_Z^2/2m_\nu=4\times10^{21}({\rm
eV}/m_\nu)\,$eV. Recently it has
been suggested that the stable end products of the ``Z-bursts''
that would thus be induced at close-by distances ($\la50\,$Mpc) from
Earth may explain the
highest energy cosmic rays~\cite{fms,weiler2} and may also
provide indirect evidence for neutrino hot dark matter. These end
products would be mostly nucleons and $\gamma-$rays with
average energies a factor of $\simeq5$ and $\simeq40$ lower, respectively,  
than the original UHE neutrino. As a consequence, if the UHE neutrino
was produced as a secondary of an accelerated proton, the energy
of the latter would have to be at least a few
$10^{22}\,$eV~\cite{fms}, making Z-bursts above GZK energies
more likely to play a role in the context of non-acceleration
scenarios. Moreover, it has
subsequently been pointed out~\cite{wax3} that Z production is
dominated by annihilation on the non-clustered massive RNB
compared to annihilation with neutrinos clustering in
the Galactic halo or in nearby galaxy clusters.
As a consequence, for a significant contribution of
neutrino annihilation to the observed EHECR flux, a new
class of neutrino sources, unrelated to UHECR sources,
seems necessary. This has been confirmed by more detailed
numerical simulations~\cite{ysl} where it has, however, also been
demonstrated that the most significant contribution could come
from annihilation on neutrino dark matter clustering in the
Local Supercluster by amounts consistent with expectations.
In the absence of any assumptions on the neutrino sources,
the minimal constraint comes from the unavoidable
production of secondary $\gamma-$rays contributing to the
diffuse flux around 10 GeV measured by EGRET: If the Z-burst
decay products are to explain EHECR, the massive neutrino
overdensity $f_\nu$ over a length scale $l_\nu$ has to
satisfy $f_\nu\ga20\,(l_\nu/5\,{\rm Mpc})^{-1}$, provided
that only neutrinos leave the source, a situation that may
arise in top-down models if the X particles decay exclusively
into neutrinos (see Fig.~\ref{fig7} below for a model involving
topological defects and Ref.~\cite{gk2} for a scenario involving
decaying superheavy relic particles). If, instead, the
total photon source luminosity is comparable to the total
neutrino luminosity, as in most models, the EGRET constraint
translates into the more stringent requirement
$f_\nu\ga10^3(l_\nu/5\,{\rm Mpc})^{-1}$. This bound can
only be relaxed if most of the EM energy is radiated in the TeV
range where the Universe is more transparent~\cite{ysl}.
Fig.~\ref{fig3} shows an example of this situation.

Furthermore, the Z-burst scenario requires sources that
are optically thick for accelerated protons with respect to
photo-pion production because otherwise the observable
proton flux below the GZK cutoff would be comparable
to the neutrino flux~\cite{slproc,wax3}. A systematic parameter study of
the required overdensity, based on analytical flux estimates,
has been performed in Ref.~\cite{bpvz1}.
Recently it has been noted that a degenerate relic neutrino
background would increase the interaction probability and
thereby make the Z-burst scenario more
promising~\cite{gelmini-kusenko}. A neutrino asymmetry of
order unity is not excluded phenomenologically~\cite{pk}
and can be created in the early Universe, for example,
through the Affleck-Dine baryogenesis mechanism~\cite{affleck-dine} or due
to neutrino oscillations. The authors of Ref.~\cite{gelmini-kusenko}
pointed out that for a neutrino mass $m_\nu\simeq0.07\,$eV,
a value suggested by the Super-Kamiokande experiment~\cite{superk},
and for sources at redshifts of a few,
the flux of secondary Z-decay products is maximal for a
RNB density parameter $\Omega_\nu\simeq0.01$. Such neutrino masses,
however, require the sources to produce neutrinos at least up
to $10^{22}\,$eV.

UHE neutrinos from the decay of pions, that are produced
by interactions of accelerated protons in astrophysical sources, must have
originated within redshifts of a few. Moreover, in most
conventional models their flux is
expected to fall off rapidly above $10^{20}\,$eV. Examples are
production in active galactic nuclei within
hadronic models~\cite{agn-nu,mannheim2,hz,protheroe2,protheroe,wb2}, and
``cosmogenic'' neutrinos from interactions of UHECR nucleons (near or 
above the GZK cutoff) with the CMB (see,
e.g.,Refs.~\cite{stecker-neut,hs}). 
The latter source is the only one that is
guaranteed to exist due to existence of UHECRs near the GZK
cutoff, but the fluxes are generally quite small. 
Therefore, interaction of these UHE neutrinos with the RNB, that could
reveal the latter's existence, can, 
if at all, be important only if the relic neutrinos have a mass
$m_\nu\ga1\,$eV~\cite{weiler1}. Due to the continuous release of
UHE neutrinos up to much higher redshifts, most top-down scenarios would imply
substantially higher fluxes that also extend to much
higher energies~\cite{bhs} (see Sect.~4.4 below). Certain features in
the UHE neutrino spectrum
predicted within such top-down scenarios, such as a change of slope for
massless neutrinos~\cite{yoshida} or a dip structure for
relic neutrino masses of order $1\,$eV~\cite{ydjs,weiler2}, have
therefore been proposed as possibly the only way to detect the RNB.
However, some of the scenarios at the high end of
neutrino flux predictions have recently been ruled out
based on constraints on the accompanying energy release into
the EM channel.

The recent claim that the Z-burst mechanism can explain
the EHECR flux even without overdensity, $f_\nu=1$,
and even allows to determine the neutrino mass
$m_\nu\sim1\,$eV~\cite{ringwald} was based on assumptions
that we characterized here as highly unrealistic:
Sources accelerating nuclei to $\ga10^{23}\,$eV
while being completely opaque to both the primaries
and the secondary photons. The energy fluence of the latter,
however, after possibly being reprocessed to lower
energies by EM cascading, must be comparable to the
neutrino energy fluence by simple isospin symmetry in
the production of charged and neutral pions.

Since in virtually all models UHE neutrinos are
created as secondaries from pion decay, i.e. as electron or muon
neutrinos, $\tau-$neutrinos can only be produced by a flavor
changing $W^\pm$ t-channel interaction with the RNB.
The flux of UHE $\tau-$neutrinos is
therefore usually expected to be substantially smaller than the
one of electron and muon neutrinos, if no neutrino oscillations
take place at these energies. However, the recent evidence
from the Superkamiokande experiment for nearly maximal
mixing between muon and $\tau-$neutrinos with $|\Delta m^2|=
|m^2_{\nu_\mu}-m^2_{\nu_\tau}|\simeq
5\times10^{-3}\,{\rm eV}^2$~\cite{superk}
would imply an oscillation length of $L_{osc}=2E/|\Delta m^2|=
2.6\times10^{-6}(E/{\rm PeV})(|\Delta m^2|/5\times10^{-3}\,{\rm eV}^2)^{-1}
\,$pc and, therefore, a rough equilibration between muon
and $\tau-$neutrino fluxes from any source at a distance
larger than $L_{osc}$~\cite{mannheim3}. Turning this around, one
sees that a source at distance $d$ emitting neutrinos of energy
$E$ is sensitive to neutrino mixing with $|\Delta m^2|=2E/d\simeq
1.3\times10^{-16}\,(E/{\rm PeV})(d/100\,{\rm Mpc})^{-1}\,
{\rm eV}^2$~\cite{pakvasa,halzen-saltzberg}. Under certain
circumstances, resonant conversion in the
potential provided by the RNB clustering in galactic halos
may also influence the flavor composition of UHE neutrinos
from extraterrestrial sources~\cite{horvat}. In addition,
such huge cosmological baselines can be sensitive probes
of neutrino decay~\cite{kmp}.

\subsection{Neutrino detection}

We now turn to a discussion of UHE neutrino interactions
with matter relevant for neutrino detection.
UHE neutrinos can be detected by detecting the muons produced 
in ordinary matter via
charged-current reactions with nucleons; see,
e.g., Refs.~\cite{fmr,gqrs,gkr} for recent discussions.
Corresponding cross sections are
calculated by folding the fundamental
standard model quark-neutrino cross section with the
distribution function of the partons in the nucleon.
These cross sections are most sensitive to the abundance of
partons of fractional momentum $x\simeq M_W^2/2m_N E$, where
$E$ is the neutrino energy. For
the relevant squared momentum transfer, $Q^2\sim M_W^2$, these
parton distribution functions have been measured down to
$x\simeq0.02$~\cite{hera}. (It has been suggested that observation
of the atmospheric neutrino flux with future neutrino telescopes
may probe parton distribution functions at much smaller $x$
currently inaccessible to colliders~\cite{ggv}).
Currently, therefore, neutrino-nucleon cross
sections for $E\ga10^{14}\,$eV can be obtained only by
extrapolating the 
parton distribution functions to lower $x$. Above
$10^{19}\,$eV, the resulting uncertainty has been estimated
to be a factor 2~\cite{gqrs}, whereas within the dynamical
radiative parton model it has been claimed to be at most
20 \%~\cite{gkr}. An intermediate estimate using the CTEQ4-DIS
distributions can roughly be parameterized by~\cite{gqrs}
\begin{equation}
  \sigma_{\nu N}(E)\simeq2.36\times10^{-32}(E/10^{19}
  \,{\rm eV})^{0.363}\,{\rm cm}^2\quad(10^{16}\,{\rm eV}\la
  E\la10^{21}\,{\rm eV})\,.\label{cccross2}
\end{equation}
Improved calculations including non-leading logarithmic
contributions in $1/x$ have recently been performed in
Ref.~\cite{kms}. The results for the neutrino-nucleon
cross section differ by less than a factor 1.5 with
Refs.~\cite{gqrs,gkr} even at $10^{21}\,$eV.

However, more recently it has been argued that the
neutrino-nucleon cross-section calculated within the Standard Model
becomes unreliable for $E\ga2\times10^{17}\,$eV: the authors 
of~\cite{dkrs} used the $O(g^2)$ expression for the elastic forward
scattering amplitude to derive via the optical theorem the bound 
$\sigma_{\nu N}\leq 9.3\times10^{-33}\,{\rm cm}^2$ for  
the total cross-section. Current parton distribution functions (pdf's) 
predict a violation of this unitarity bound above
$E\ga 2\times10^{17}\,$eV. The authors of~\cite{dkrs} argue that the
large $O(g^4)$ corrections to the forward scattering amplitude necessary 
to restore  unitarity signal a breakdown of electroweak perturbation
theory. Alternatively, large changes in the evolution of
the parton distribution functions have to set in 
soon after the kinematical range probed by HERA.
A large $O(g^4)$ correction to the forward scattering amplitude is, 
however, not surprising because the $O(g^2)$ amplitude contains no resonant 
contribution and is real. In particular, the total cross-section is
therefore not only 
bounded by a constant but zero at $O(g^2)$, and the imaginary part of
the box diagram of $O(g^4)$ is the {\em first\/} contribution to the total
neutrino-nucleon  cross-section~\cite{kp1}.

Interestingly, it has been shown that the increasing
target mass provided by the Earth for increasing zenith angles
below the horizontal implies that the rate of
up-going air showers in UHECR detectors does not decrease
with decreasing neutrino-nucleon cross section but may even
increase~\cite{upgoing}.
Thus, cross sections smaller than Eq.~(\ref{cccross2}) do
not lead to reduced event rates in UHECR detectors and can
be measured from the angular distribution of events.
UHECR and neutrino experiments can thus contribute to
measure cross sections at energies inaccessible in accelerator
experiments!

Neutral-current neutrino-nucleon cross sections are
expected to be a factor 2-3 smaller than charged-current cross
sections at UHE and interactions with electrons only play a
role at the Glashow resonance, $\bar\nu_e e\to W$, at
$E=6.3\times10^{15}\,$eV. Furthermore, cross sections of
neutrinos and anti-neutrinos are basically identical at UHE.
Radiative corrections influence the total cross section
negligibly compared to the parton distribution uncertainties,
but may lead to an increase of the average inelasticity in
the outgoing lepton from $\simeq0.19$ to $\simeq0.24$ at
$E\sim10^{20}\,$eV~\cite{sigl}, although this would
probably hardly influence the shower character.

Neutrinos propagating through the Earth start to be attenuated
above $\simeq100\,$TeV due to the increasing Standard Model
cross section as indicated by Eq.~(\ref{cccross2}).
Detailed integrations of the relevant transport equations for
muon neutrinos above a TeV have been presented in Ref.~\cite{kms},
and, for a general cold medium, in Ref.~\cite{np}.
In contrast, $\tau-$neutrinos with energy up to $\simeq100\,$PeV 
can penetrate the Earth due to their
regeneration from $\tau$ decays~\cite{halzen-saltzberg}.
As a result, a primary UHE $\tau-$neutrino beam propagating
through the Earth would cascade down below $\simeq100\,$TeV
and in a neutrino telescope could give rise to a higher total
rate of upgoing events as compared to downgoing events for
the same beam arriving from above the horizon. As mentioned
above, a primary $\tau-$neutrino beam could arise even in
scenarios based on pion decay, if $\nu_\mu-\nu_\tau$ mixing
occurs with the parameters suggested by the Super-Kamiokande
results~\cite{mannheim3}. In the PeV range, $\tau-$neutrinos
can produce characteristec "double-bang" events where the first
bang would be due to the charged-current production by the
$\tau-$neutrino of a $\tau$ whose decay at a typical distance
$\simeq$ 100\ m would produce the second bang~\cite{pakvasa}. These
effects have also been suggested as an independent astrophysical
test of the neutrino oscillation hypothesis. In addition,
isotropic neutrino fluxes in the energy range between 10 TeV and
10 PeV have been suggested as probes of the Earth's density
profile, whereby neutrino telescopes could be used for
neutrino absorption tomography~\cite{jrf}.

\subsection{New Interactions}

It has been suggested that the neutrino-nucleon
cross section, $\sigma_{\nu N}$, can be enhanced by new
physics beyond the electroweak scale in the center of
mass (CM) frame, or above about a PeV in the nucleon rest
frame~\cite{bhfpt,sigl1,jain}.
Neutrino induced air showers may therefore rather directly
probe new physics beyond the electroweak scale.

The lowest partial wave contribution to the cross section of a
point-like particle is constrained by unitarity to be not
much larger than a typical electroweak cross section~\cite{bhg}.
However, at least two major possibilities allowing considerably
larger cross sections have been discussed in the literature
for which unitarity bounds need not be violated. In the first, a broken SU(3)
gauge symmetry dual to the unbroken SU(3) color gauge group
of strong interaction is introduced as the ``generation symmetry'' such
that the three generations of leptons and quarks represent the quantum
numbers of this generation symmetry. In this scheme, neutrinos can have
close to strong interaction cross sections with quarks.
In addition, neutrinos can
interact coherently with all partons in the nucleon, resulting in
an effective cross section comparable to the geometrical
nucleon cross section.  This model lends itself to experimental
verification through shower development altitude
statistics~\cite{bhfpt}.

The second possibility consists of a large increase
in the number of degrees of freedom above the electroweak 
scale~\cite{kovesi-domokos}. A specific implementation
of this idea is given in theories with $n$ additional large
compact dimensions and a quantum gravity scale $M_{4+n}\sim\,$TeV
that has recently received much attention in the
literature~\cite{tev-qg} because it provides an alternative
solution (i.e., without supersymmetry) to the hierarchy problem
in grand unifications of gauge interactions.
The cross sections within such scenarios have not been
calculated from first principles yet. Within the field
theory approximation which should hold for squared CM
energies $s\la M_{4+n}^2$, the spin 2 character
of the graviton predicts $\sigma_g\sim s^2/M_{4+n}^6$~\cite{ns}.
For $s\gg M_{4+n}^2$, several arguments based on unitarity
within field theory have been put forward. The emission
of massive Kaluza-Klein (KK) graviton modes associated
with the increased phase space due to the extra dimensions
leads to the rough estimate (derived for $n=2$)~\cite{ns}
\begin{equation}
  \sigma_{g}\simeq\frac{4\pi s}{M^4_{4+n}}\simeq
  10^{-27}\left(\frac{M_{4+n}}{{\rm TeV}}\right)^{-4}
  \left(\frac{E}{10^{20}\,{\rm eV}}\right)\,{\rm cm}^2\,,
  \label{sigma_graviton}
\end{equation}
where in the last expression we specified to a neutrino
of energy $E$ hitting a nucleon at rest. A more detailed
calculation taking into account scattering on individual
partons leads to similar orders of magnitude~\cite{jain}.
Note that a neutrino would typically start to interact in the atmosphere
for $\sigma_{\nu N}\ga10^{-27}\,{\rm cm}^2$, i.e. in the
case of Eq.~(\ref{sigma_graviton}) for $E\ga10^{20}\,$eV, assuming
$M_{4+n}\simeq1\,$TeV. For cross sections such large the neutrino
therefore becomes a primary candidate for the
observed EHECR events. However, since in a neutral current
interaction the neutrino transfers only about $10\%$ of its energy to
the shower, the cross section probably has to be at least a
few $10^{-26}\,{\rm cm}^2$ to be consistent with observed
showers which start within the first $50\,{\rm g}\,{\rm cm}^{-2}$
of the atmosphere~\cite{kp,agmprs}.
A specific signature of this scenario
would be the absence of any events above the energy where
$\sigma_g$ grows beyond $\simeq10^{-27}\,{\rm cm}^2$ in
neutrino telescopes based on ice or water as detector
medium~\cite{nutel}, and a hardening of the spectrum above this energy
in atmospheric detectors such as the Pierre Auger
Project~\cite{auger} and the proposed space based AirWatch type
detectors~\cite{owl,euso,airwatch}.
Furthermore, according to Eq.~(\ref{sigma_graviton}),
the average atmospheric column depth of the first interaction
point of neutrino induced EAS in this scenario is predicted
to depend linearly on energy. This should be easy to distinguish
from the logarithmic scaling expected
for nucleons, nuclei, and $\gamma-$rays. To test
such scalings one can, for example, take advantage of the
fact that the atmosphere provides a detector medium whose
column depth increases from $\sim1000\,{\rm g}/{\rm cm}^2$
towards the zenith to $\sim36000\,{\rm g}/{\rm cm}^2$
towards horizontal arrival directions. This probes
cross sections in the range $\sim10^{-29}-10^{-27}\,{\rm cm}^2$.
Due to the increased column depth, water/ice detectors would probe
cross sections in the range
$\sim10^{-31}-10^{-29}\,{\rm cm}^2$~\cite{tol} which could be
relevant for TeV scale gravity models~\cite{bhhk}.

\begin{figure}
\begin{center}
\centerline{\hbox{\psfig{figure=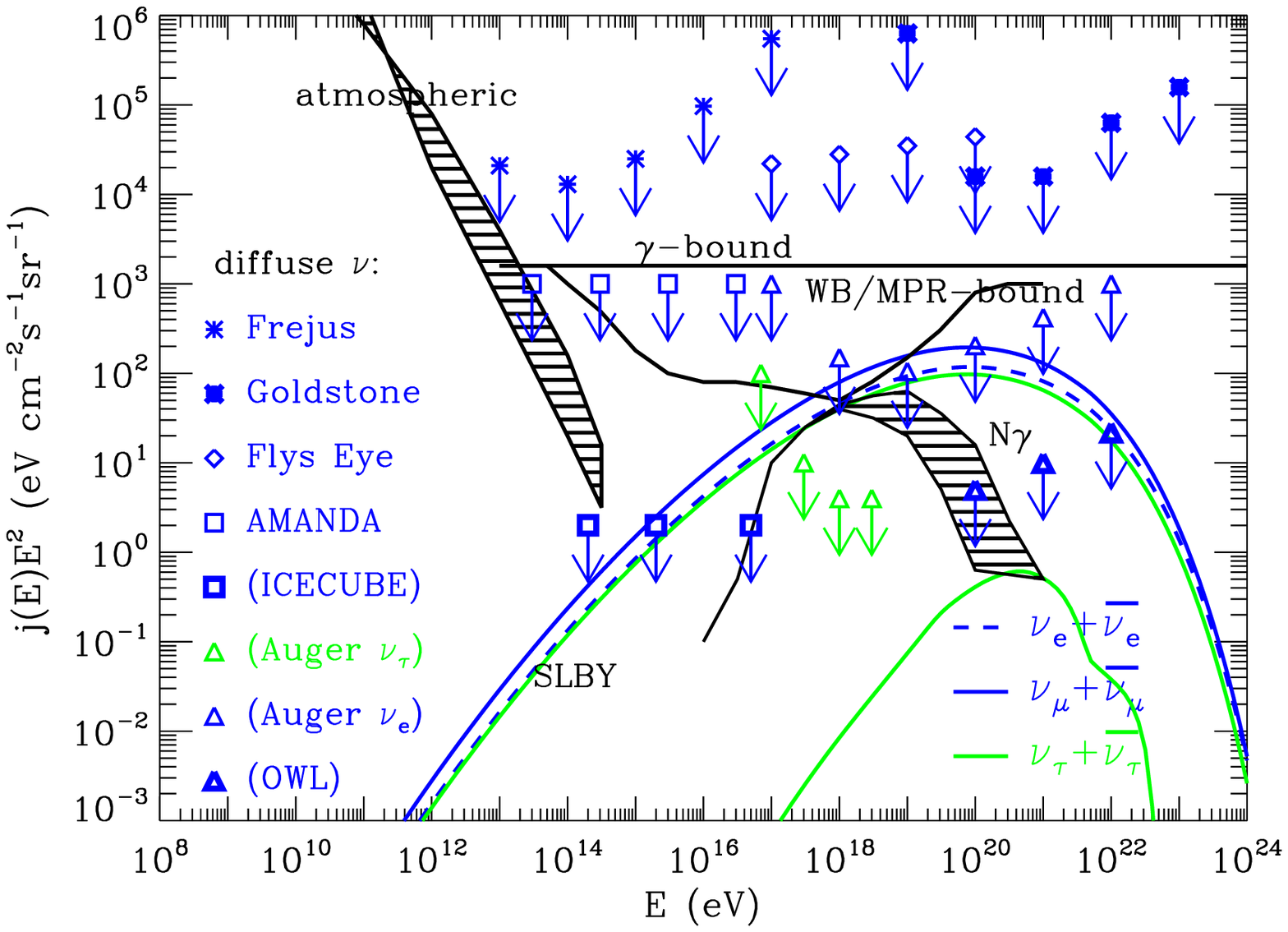,clip=true,width=0.9\textwidth}}}
\end{center}
\caption[...]{Various neutrino flux predictions and experimental upper
limits or projected sensitivities. Shown
are upper limits from the Frejus underground
detector~\cite{frejus}, the Fly's Eye experiment~\cite{baltrusaitis}, the
Goldstone radio telescope~\cite{goldstone}, and the
Antarctic Muon and Neutrino Detector Array (AMANDA) neutrino
telescope~\cite{amanda}, as well as projected neutrino flux
sensitivities of ICECUBE, the planned kilometer scale extension
of AMANDA~\cite{icecube}, the Pierre Auger Project~\cite{auger-neut}
(for electron and tau neutrinos separately) and the proposed
space based OWL~\cite{owl} concept. Neutrino fluxes
are shown for the atmospheric neutrino background~\cite{lipari} (hatched
region marked ``atmospheric''), for EHECR interactions with the CMB~\cite{pj}
(``$N\gamma$'', dashed range indicating typical uncertainties
for moderate source evolution), and for the ``top-down'' model
(marked ``SLBY'') corresponding to Fig.~\ref{fig6} of Sect.~4
below, where EHECR and neutrinos are produced by decay of
superheavy relics.
The top-down fluxes are shown for electron-, muon, and tau-neutrinos
separately, assuming no (lower $\nu_\tau$-curve) and maximal
$\nu_\mu-\nu_\tau$ mixing (upper $\nu_\tau$-curve, which would then
equal the $\nu_\mu$-flux), respectively. The Waxman-Bahcall bound
in the version of Mannheim, Protheroe, and Rachen~\cite{mpr}
(``WB/MPR-bound'') for sources optically
thin for the proton primaries, and the $\gamma-$ray bound
(``$\gamma-$bound'') are also shown.}
\label{fig4}
\end{figure}

From a perturbative point of view within string theory, $\sigma_{\nu N}$
can be estimated as follows: Individual amplitudes are expected
to be suppressed exponentially above the string scale $M_s$
which again for simplicity we assume here to be comparable to $M_{4+n}$.
This can be interpreted as a result of the finite spatial
extension of the string states.
In this case, the neutrino nucleon cross section would
be dominated by interactions with the partons carrying
a momentum fraction $x\sim M_s^2/s$, leading to~\cite{kp}
\begin{eqnarray}
  \sigma_{\nu N}&\simeq&\frac{4\pi}{M_s^2}\ln(s/M_s^2)
  (s/M_s^2)^{0.363}\simeq
  6\times10^{-29}\left(\frac{M_s}{{\rm TeV}}\right)^{-4.726}
  \left(\frac{E}{10^{20}\,{\rm eV}}\right)^{0.363}\nonumber\\
  &&\times\left[1+0.08\ln\left(\frac{E}{10^{20}\,{\rm eV}}\right)
  -0.16\ln\left(\frac{M_s}{{\rm TeV}}\right)\right]^2\,{\rm cm}^2
  \label{sigma_nuN}
\end{eqnarray}
This is probably too small to make neutrinos primary candidates
for the highest energy showers observed, given the
fact that complementary constraints from accelerator
experiments result in $M_s\ga 1\,$TeV~\cite{coll-extra-dim}.
On the other hand, general arguments on the production of
``string balls'' or small black holes from two point particles
represented by light strings~\cite{de} leads to the asymptotic
scaling $\sigma\simeq(s/M_s^2)^{1/(n+1)}$ for $s\ga M_s^2$ for
the fundamental neutrino-parton cross section. This could
lead to values for $\sigma_{\nu N}$ larger than
Eq.~(\ref{sigma_nuN})~\cite{fs}.
Some other recent work seems to imply that in TeV string models
cross sections, if not sufficient to make neutrinos UHECR primary
candidates, could at least be significantly larger than
Standard Model cross sections~\cite{cim}.
Thus, an experimental detection of the signatures discussed
in this section could lead to constraints on some
string-inspired models of extra dimensions.

There are, however, severe astrophysical and cosmological
constraints on $M_{4+n}$ which result from limiting the emission
of bulk gravitons into the extra dimensions. 
The strongest constraints in this regard come from the
production due to nucleon-nucleon
bremsstrahlung in type II supernovae~\cite{astro-extra-dim}
and their subsequent decay into a diffuse background of
$\gamma-$rays in the MeV range~\cite{hr}. The latter read
$M_6\ga84\,$TeV, $M_7\ga7\,$TeV, for $n=2,3$, respectively,
and, therefore, $n\geq5$ is required if neutrino primaries
are to serve as a primary candidate for the EHECR events observed
above $10^{20}\,$eV. This assumes a toroidal geometry of the
extra dimensions with equal radii given by
\begin{equation}
  r_n\simeq M^{-1}_{4+n}\left(\frac{M_{\rm Pl}}{M_{4+n}}\right)^{2/n}
  \simeq2\times10^{-17}\left(\frac{{\rm TeV}}{M_{4+n}}\right)
  \left(\frac{M_{\rm Pl}}{M_{4+n}}\right)^{2/n}\,{\rm cm}
  \,,\label{rextra}
\end{equation}
where $M_{\rm Pl}$ denotes the Planck mass. 
The above lower bounds on $M_{4+n}$ thus translate into the corresponding
upper bounds $r_2\la0.9\times10^{-4}\,$mm, $r_3\la0.19\times10^{-6}\,$mm,
respectively. Still stronger but somewhat more model
dependent bounds result from the production of KK modes during
the reheating phase after inflation. For example, a bound
$M_{6}\ga500\,$TeV has been reported~\cite{hannestad} based
on the contribution to the diffuse $\gamma-$ray background
in the 100 MeV region. We note, however, that all astrophysical
and cosmological bounds are changed in more complicated geometries
of extra dimensions~\cite{dienes}.

The neutrino primary hypothesis of EHECR together with other astrophysical
and cosmological constraints thus provides an interesting testing
ground for theories involving large compact extra dimensions representing 
one possible kind of physics beyond the Standard Model. 
In this context, we mention that in theories with large compact extra
dimensions mentioned above, Newton's law of gravity is expected to be
modified at distances smaller than the length scale given by
Eq.~(\ref{rextra}). Indeed, there are laboratory 
experiments measuring gravitational interaction at small
distances (for a recent review of such experiments see
Ref.~\cite{lcp}), which also probe these theories. Thus, future EHECR
experiments and gravitational experiments in the laboratory together 
have the potential of providing rather strong tests of these theories. 
These tests would be complementary to constraints
from collider experiments~\cite{coll-extra-dim}.

Independent of theoretical arguments, the EHECR data
can be used to put constraints on cross sections
satisfying $\sigma_{\nu N}(E\ga10^{19}\,{\rm eV})
\la10^{-27}\,{\rm cm}^2$. Particles with such cross
sections would give rise to horizontal air showers.
The Fly's Eye experiment established an upper limit
on horizontal air showers~\cite{baltrusaitis}. The non-observation of
the neutrino flux expected from pions produced by
EHECRs interacting with the CMB the results in the
limit~\cite{mr,tol}
\begin{eqnarray}
  \sigma_{\nu N}(10^{17}\,{\rm eV})&\la&1\times10^{-29}
  /{\bar y}^{1/2}\,{\rm cm}^2\nonumber\\
  \sigma_{\nu N}(10^{18}\,{\rm eV})&\la&8\times10^{-30}
  /{\bar y}^{1/2}\,{\rm cm}^2\nonumber\\
  \sigma_{\nu N}(10^{19}\,{\rm eV})&\la&5\times10^{-29}
  /{\bar y}^{1/2}\,{\rm cm}^2\,,\label{crosslim1}
\end{eqnarray}
where ${\bar y}$ is the average energy fraction of the neutrino
deposited into the shower (${\bar y}=1$ for charged current
reactions and ${\bar y}\simeq0.1$ for neutral current reactions).
Neutrino fluxes predicted in various scenarios are shown in Fig.~\ref{fig4}.
The projected sensitivity of future experiments such as
the Pierre Auger Observatories and the AirWatch type satellite
projects indicate that the cross section limits Eq.~(\ref{crosslim1})
could be improved by up to four orders of magnitude,
corresponding to one order of magnitude in $M_s$ or
$M_{4+n}$. This would close the window between cross sections
allowing horizontal air showers, $\sigma_{\nu N}(E\ga10^{19}\,{\rm eV})
\la10^{-27}\,{\rm cm}^2$, and the Standard Model value
Eq.~(\ref{cccross2}).

We note in this context that, only assuming
$3+1$ dimensional field theory, consistency of the UHE $\nu N$ cross
section with data at electroweak energies does not lead to very
stringent constraints: Relating the cross section to the
$\nu N$ elastic amplitude in a model independent way yields~\cite{gw}
\begin{equation}
\sigma(E)\la3\times10^{-24}\left(\frac{E}{10^{19}\,{\rm eV}}
\right)\,{\rm cm}^2\,.\label{gwbound}
\end{equation}
However, it has been argued~\cite{dkrs} (see discussion
above), cross sections $\sigma_{\nu N}\ga9.3\times10^{-33}\,{\rm cm}^2$
could signal a breakdown of perturbation theory.

In the context of conventional astrophysical sources, the relevant UHE
neutrino primaries could, of course, only be produced as secondaries
in interactions with matter or with low energy photons of protons or
nuclei accelerated to energies of at least $10^{21}\,$eV.
This implies strong requirements on the possible sources.

In addition, neutrino primaries with new interactions would predict
a significant correlation of UHECR arrival directions with high
redshift objects. Indeed, various claims recently occured in the
literature for significant
angular correlations of UHECRs with certain astrophysical objects at
distances too large for the primaries to be nucleons, nuclei,
or $\gamma-$rays. Farrar and Biermann reported a possible correlation between
the arrival direction of the five highest energy CR events and compact radio
quasars at redshifts between 0.3 and 2.2~\cite{fb}
Undoubtedly, with the present amount of data the interpretation
of such evidence for a correlation remains somewhat subjective,
as is demonstrated by the criticism of the statistical analysis
in Ref.~\cite{fb} by Hoffman~\cite{hoffman} and the reply by Farrar
and Biermann~\cite{fb-reply}). Also, a new analysis with the somewhat
larger data set now available did not support such
correlations~\cite{star}. This is currently disputed
since another group claims to have found a correlation
on the 99.9\% confidence level~\cite{virmani}. Most recently, a
correlation between UHECRs of energy $E\ga4\times10^{19}\,$eV
and BL Lacertae objects at redshifts $z>0.1$ was claimed~\cite{tt}.
None of these claims are convincing yet but confirmation
or refutation should be possible within the next few years
by the new experiments. Clearly, a confirmation of one of these
correlations would be exciting as it would probably imply new physics
such as neutrinos with new interactions, as yet unknown, e.g.
supersymmetric, particles, or even more radical propositions
such as violation of Lorentz invariance (for a discussion
of these latter possibilities see, e.g., Refs.~\cite{bs-rev}.

\section{Top-Down Scenarios}

\subsection{The Main Idea}

As mentioned in the introduction, all top-down scenarios
involve the decay of X particles of mass close to the GUT scale
which can basically be produced in two ways: If they are very
short lived, as usually expected in many GUTs, they have to be
produced continuously. The only way this can be achieved is
by emission from topological defects left over from cosmological
phase transitions that may have occurred in the early Universe at
temperatures close to the GUT scale, possibly during reheating
after inflation. Topological defects
necessarily occur between regions that are causally disconnected, such
that the orientation of the order parameter
associated with the phase transition can not be communicated
between these regions and consequently will adopt different
values. Examples are cosmic strings (similar to vortices in superfluid
helium), magnetic monopoles, and domain walls (similar to Bloch
walls separating regions of different magnetization in a ferromagnet).
The defect density is consequently given by the particle horizon
in the early Universe and their formation can even be studied
in solid state experiments where the expansion rate of the Universe
corresponds to the quenching speed with which the phase
transition is induced~\cite{vachaspati}. The defects are
topologically stable, but
in the cosmological case time dependent motion
leads to the emission of particles with a mass comparable to the
temperature at which the phase transition took place. The
associated phase transition can also occur during reheating
after inflation.

Alternatively, instead of being released from topological
defects, X particles
may have been produced directly in the early Universe and,
due to some unknown symmetries, have a very
long lifetime comparable to the age of the Universe.
In contrast to Weakly-Interacting Massive Particles (WIMPS)
below a few hundred TeV which are the usual dark matter
candidates motivated by, for example, supersymmetry and can
be produced by thermal freeze out, such superheavy X particles
have to be produced non-thermally.
Several such mechanisms operating in the
post-inflationary epoch in the early Universe have been studied. They
include gravitational production through the effect of the expansion of
the background metric on the vacuum quantum fluctuations of the
X particle field, or creation during reheating at
the end of inflation if the X particle field couples to the inflaton
field. The latter case can be divided into
three subcases, namely ``incoherent'' production with an
abundance proportional to the X particle annihilation cross section,
non-adiabatic production in broad parametric resonances with
the oscillating inflaton field during preheating (analogous
to energy transfer in a system of coupled pendula), and creation
in bubble wall collisions if inflation is completed by a first
order phase transition. In all these cases, such particles,
also called ``WIMPZILLAs'', would contribute to the dark matter
and their decays could still contribute to UHE CR fluxes today,
with an anisotropy pattern that reflects the dark matter
distribution in the halo of our Galaxy.

It is interesting to note that one of the prime motivations
of the inflationary paradigm was to dilute excessive production
of ``dangerous relics'' such as topological defects and
superheavy stable particles. However, such objects can be
produced right after inflation during reheating
in cosmologically interesting abundances, and with a mass scale
roughly given by the inflationary scale which in turn
is fixed by the CMB anisotropies to
$\sim10^{13}\,$GeV~\cite{kuz-tak}. The reader will realize that
this mass scale is somewhat above the highest energies
observed in CRs, which implies that the decay products of
these primordial relics could well have something to do with
EHECRs which in turn can probe such scenarios!

For dimensional reasons the spatially averaged X particle
injection rate can only
depend on the mass scale $m_X$ and on cosmic time $t$ in the
combination
\begin{equation}
  \dot n_X(t)=\kappa m_X^p t^{-4+p}\,,\label{dotnx}
\end{equation}
where $\kappa$ and $p$ are dimensionless constants whose
value depend on the specific top-down scenario~\cite{bhs},
For example, the case $p=1$ is representative of scenarios
involving release of X particles from topological defects,
such as ordinary cosmic
strings~\cite{br}, necklaces~\cite{bv} and magnetic 
monopoles~\cite{bs}. This can be easily seen as follows:
The energy density $\rho_s$ in a network of defects has to scale
roughly as the critical density, $\rho_s\propto\rho_{\rm crit}\propto
t^{-2}$, where $t$ is cosmic time, otherwise the defects
would either start to overclose the Universe, or end up
having a negligible contribution to the total energy
density. In order to maintain this scaling, the defect
network has to release energy with a rate given by
$\dot\rho_s=-a\rho_s/t\propto t^{-3}$, where $a=1$ in
the radiation dominated aera, and $a=2/3$ during matter
domination. If most of this energy goes into emission
of X particles, then typically $\kappa\sim{\cal O}(1)$.
In the flux calculations for TD models presented in this
paper, it was assumed that the X particles are nonrelativistic at decay.

The X particles could be gauge bosons, Higgs bosons, superheavy fermions,
etc.~depending on the specific GUT. They would have
a mass $m_X$ comparable to the symmetry breaking scale and would
decay into leptons and/or quarks of roughly
comparable energy. The quarks interact strongly and 
hadronize into nucleons ($N$s) and pions, the latter
decaying in turn into $\gamma$-rays, electrons, and neutrinos. 
Given the X particle production rate, $dn_X/dt$, the effective
injection spectrum of particle species $a$ ($a=\gamma,N,e^\pm,\nu$) 
via the hadronic channel can be
written as $(dn_X/dt)(2/m_X)(dN_a/dx)$,
where $x \equiv 2E/m_X$, and $dN_a/dx$ is the relevant
fragmentation function (FF).

We adopt the Local Parton Hadron Duality (LPHD) approximation~\cite{detal}
according to which the total
hadronic FF, $dN_h/dx$, is taken to be proportional to the spectrum
of the partons (quarks/gluons) in the parton cascade (which is initiated
by the quark through perturbative QCD processes) after evolving the parton
cascade to a stage where the typical transverse momentum transfer in the
QCD cascading processes has come down to $\sim R^{-1}\sim$ few hundred 
MeV, where $R$ is a typical hadron size. The parton spectrum is obtained
from solutions of the standard QCD evolution equations in modified leading
logarithmic approximation (MLLA) which provides good fits to accelerator
data at LEP energies~\cite{detal}. We will specifically use a recently
suggested generalization of the MLLA spectrum that includes the effects 
of supersymmetry~\cite{bk}. Within the LPHD hypothesis, the pions
and nucleons after hadronization have essentially the same spectrum. 
The LPHD does not, however, fix the relative abundance of pions and
nucleons after hadronization. Motivated by accelerator data, we assume
the nucleon content $f_N$ of the hadrons to be in the range
3 to 10\%, independent of energy, and the rest pions distributed
equally among the three charge states. For $x\la0.1$ this
ratio as well as the spectra are roughly consistent with recent more
detailed Monte Carlo simulations~\cite{BirSar,berek}. This is also
the range relevant for comparison with observations
as long as $m_X\gg10^{12}\,$GeV, which covers the range considered
in the present paper. Resulting normalization
ambiguities~\cite{BirSar} are of factors of a few and thus
subdominant in light of many other uncertainties at the current
stage of these scenarios.
The standard pion decay spectra then give the injection spectra
of $\gamma$-rays, electrons, and neutrinos. For more details
concerning uncertainties in the X particle decay spectra
see Ref.~\cite{slby}.

\subsection{Numerical Simulations}
The $\gamma$-rays and electrons produced by  X particle decay
initiate  electromagnetic
(EM) cascades on low energy radiation fields such as the
CMB. The high energy photons undergo electron-positron pair
production (PP; $\gamma \gamma_b \rightarrow e^- e^+$), and 
at energies below $\sim 10^{14}$ eV they interact mainly with 
the universal infrared and optical (IR/O) backgrounds, while above 
$\sim 100$ EeV  they interact mainly with the universal radio background (URB).
In the Klein-Nishina regime, where the CM energy is
large compared to the electron mass, one of the outgoing particles usually
carries most of the initial energy. This ``leading''
electron (positron) in turn can transfer almost all of its energy to
a background photon via inverse
Compton scattering (ICS; $e \gamma_b \rightarrow e^\prime\gamma$).
EM cascades are driven by this cycle of PP and ICS.
The energy degradation of the ``leading'' particle in this cycle
is slow, whereas the total number of particles grows
exponentially with time. This makes a standard Monte Carlo
treatment difficult. Implicit numerical schemes have therefore been
used to solve the relevant kinetic 
equations. A detailed account of the transport equation approach
used in the calculations whose results are presented in this
contribution can be found in Ref.~\cite{Lee}. All
EM interactions that influence the $\gamma$-ray spectrum in the energy range
$10^8\,{\rm eV} < E < 10^{25}\,$eV, namely PP, ICS, triplet pair
production (TPP; $e \gamma_b
\rightarrow e e^- e^+$), and double pair production (DPP, $\gamma \gamma_b
\rightarrow e^-e^+e^-e^+$), as well as synchrotron losses
of electrons in the large scale extragalactic magnetic field
(EGMF), are included.

Similarly to photons, UHE neutrinos give rise to neutrino
cascades in the primordial neutrino background via exchange
of W and Z bosons~\cite{Zburst,ydjs}. Besides the secondary
neutrinos which drive the neutrino cascade, the W and Z decay products
include charged leptons and quarks which in turn feed into the
EM and hadronic channels. Neutrino interactions become
especially significant if the relic neutrinos have masses $m_\nu$
in the eV range and thus constitute hot dark matter, because
the Z boson resonance then occurs at an UHE neutrino energy
$E_{\rm res}=4\times10^{21}({\rm eV}/m_\nu)$ eV. In fact, this has been
proposed as a significant source of EHECRs~\cite{weiler2,ysl}.
Motivated by recent experimental evidence for neutrino mass
we assume the neutrino masses to be
$m_{\nu_e}=0.1\,$eV, $m_{\nu_\mu}=m_{\nu_\tau}=1\,$eV
and implemented the relevant W boson interactions in the
t-channel and the Z boson exchange via t- and s-channel. Hot dark matter
is also expected to cluster, potentially increasing secondary
$\gamma$-ray and nucleon production~\cite{weiler2,ysl}. This influences
mostly scenarios where X decays into neutrinos only. We
parametrize massive neutrino clustering by a length scale $l_\nu$
and an overdensity $f_\nu$ over the average density $\bar{n_\nu}$.
The Fermi distribution with a velocity dispersion $v$ yields
$f_\nu\la v^3 m_\nu^3/(2\pi)^{3/2}/\bar{n_\nu}\simeq
330\,(v/500\,{\rm km}\,{\rm sec}^{-1})^3\,
(m_\nu/{\rm eV})^3$~\cite{peebles}. Therefore, values of
$l_\nu\simeq$ few Mpc and $f_\nu\simeq20$ are conceivable
on the local Supercluster scale~\cite{ysl}.

The relevant nucleon interactions implemented are
pair production by protons ($p\gamma_b\rightarrow p e^- e^+$),
photoproduction of single or multiple pions ($N\gamma_b \rightarrow N
\;n\pi$, $n\geq1$), and neutron decay.
In TD scenarios, the particle injection spectrum is generally dominated
by the ``primary'' $\gamma$-rays and neutrinos over nucleons. These
primary $\gamma$-rays and neutrinos are produced by the decay of
the primary pions resulting from the hadronization of quarks that come
from the decay of the X particles. The contribution of secondary
$\gamma$-rays, electrons, and neutrinos from decaying pions that are
subsequently produced by the interactions of nucleons with
the CMB, is in general negligible compared to that of the primary
particles; we nevertheless include the contribution of the
secondary particles in our code.

In principle, new interactions such as the ones involving
a TeV quantum gravity scale can not only modify the interactions
of primary particles in the detector, as discussed in Sect.~3.3,
but also their propagation. However, $\gamma-$rays and
nuclei interact mostly with the CMB and IR for which the
CM energy is at most
$\simeq30(E/10^{14}\,{\rm GeV})^{1/2}\,$GeV. At such energies
the new interactions are much weaker than the dominating
electromagnetic and strong interactions. It has been
suggested recently~\cite{neutrino_int} that new interactions
may notably influence UHE neutrino propagation. However,
for neutrinos of mass $m_\nu$, the CM energy is
$\simeq100(E/10^{14}\,{\rm GeV})^{1/2}(m_\nu/0.1\,{\rm eV})^{1/2}\,$GeV,
and therefore, UHE neutrino propagation would only be
significantly modified for neutrino masses significantly
larger than $0.1\,$eV. We will therefore ignore this possibility
here.

We assume a flat Universe with no cosmological constant,
and a Hubble constant of $h=0.65$ in units of
$100\;{\rm km}\;{\rm sec}^{-1}{\rm Mpc}^{-1}$ throughout.
The numerical calculations follow
{\it all} produced particles in the EM, hadronic,
and neutrino channel, whereas the often-used continuous energy loss (CEL)
approximation (e.g., \cite{ABS}) follows
only the leading cascade particles. The CEL approximation
can significantly underestimate the cascade flux at lower energies.

The two major uncertainties in the particle transport are the
intensity and spectrum of the URB for which there exists only
an estimate above a few MHz frequency~\cite{Clark}, and the average value
of the EGMF. To bracket these uncertainties, simulations
have been performed for the observational URB estimate from
Ref.~\cite{Clark} that
has a low-frequency cutoff at 2 MHz (``minimal''), and the medium
and maximal theoretical estimates
from Ref.~\cite{pb}, as well as for EGMFs between zero
and $10^{-9}$ G, the latter motivated by limits from
Faraday rotation measurements~\cite{rkb}. A strong URB tends
to suppress the UHE $\gamma$-ray flux by direct absorption
whereas a strong EGMF blocks EM cascading (which otherwise develops
efficiently especially in a low URB) by synchrotron cooling
of the electrons. For the IR/O background we used the most
recent data~\cite{irb}.

\subsection{Results: $\gamma-$ray and Nucleon Fluxes}

\begin{figure}[htb]
\centerline{\hbox{\psfig{figure=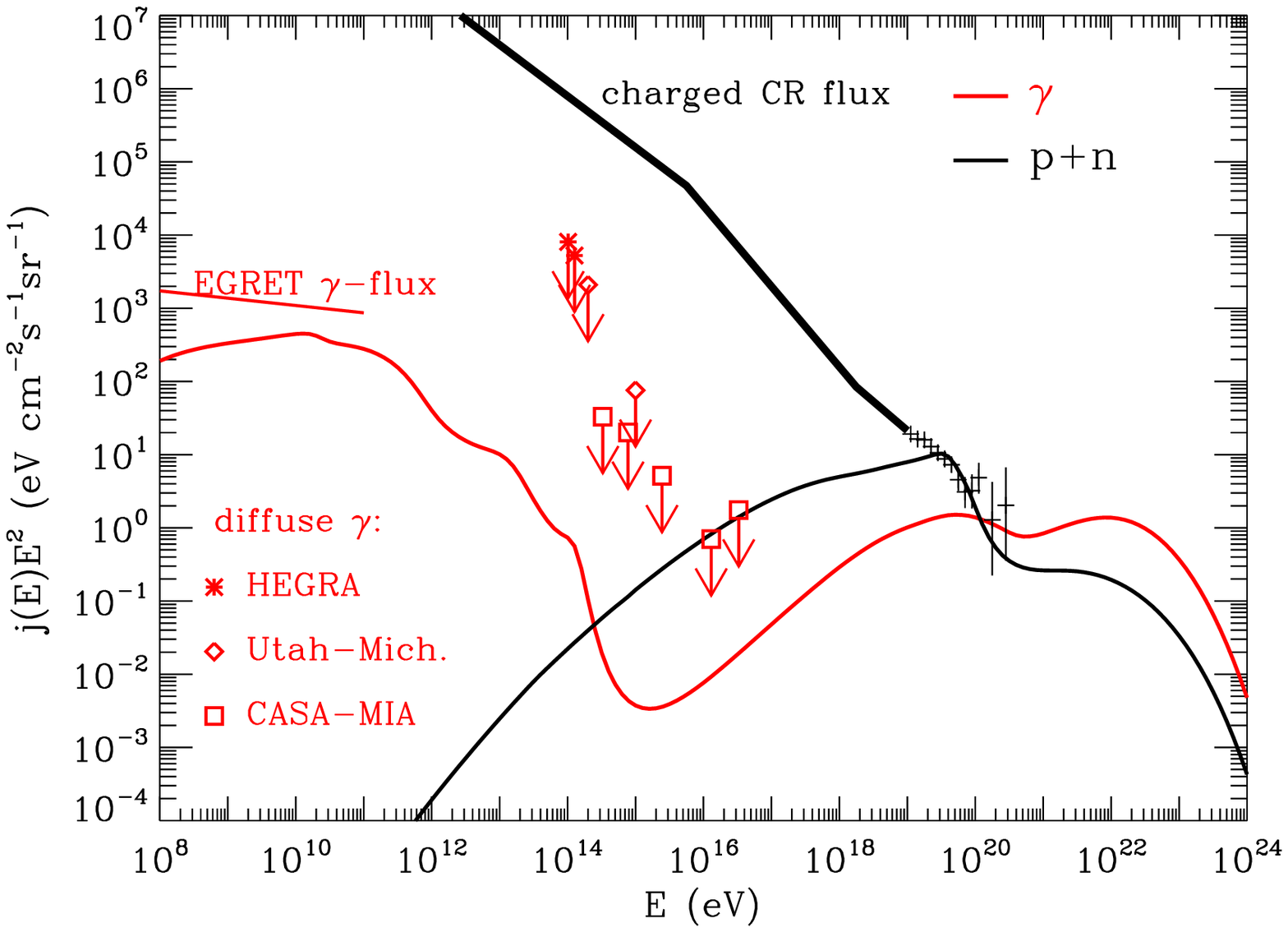,clip=true,width=0.9\textwidth}}}
\caption[...]{Predictions for the differential fluxes of
$\gamma-$rays (solid line) and protons and neutrons (dotted
line) in a TD model characterized by $p=1$, $m_X = 10^{16}\,$GeV,
and the decay mode $X\to q+q$, assuming the supersymmetric modification of
the fragmentation function~\cite{bk}, with a fraction of about
10\% nucleons. The defects have been assumed to be homogeneously
distributed. The calculation used the code described in Ref.~\cite{slby}
and assumed an intermediate URB estimate from Ref.~\cite{pb} and
an EGMF $\ll10^{-11}\,$G. Remaining line key as in Fig.~\ref{fig3}.
\label{fig5}}
\end{figure}

\begin{figure}[htb]
\centerline{\hbox{\psfig{figure=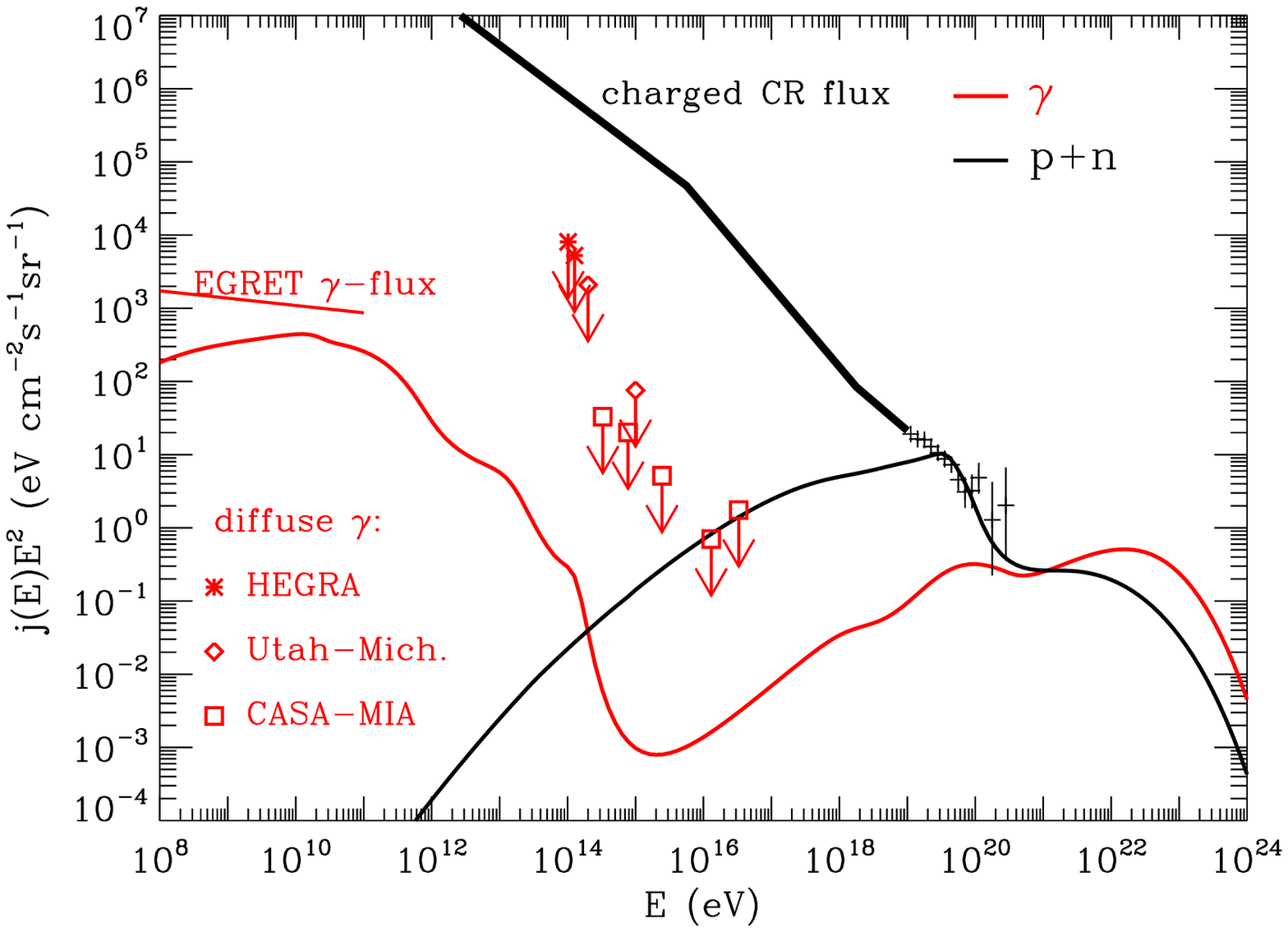,clip=true,width=0.9\textwidth}}}
\caption[...]{Same as Fig.~\ref{fig5}, but for an EGMF of
$10^{-9}\,$G.
\label{fig6}}
\end{figure}

Fig.~\ref{fig5} shows results from Ref.~\cite{slby} for the
time averaged $\gamma-$ray and nucleon
fluxes in a typical TD scenario, assuming no
EGMF, along with current observational constraints on
the $\gamma-$ray flux. The spectrum was optimally normalized 
to allow for an explanation of the observed EHECR
events, assuming their consistency with a nucleon or
$\gamma-$ray primary. The flux below $\la2\times10^{19}\,$eV
is presumably due to conventional
acceleration in astrophysical sources and was not fit. Similar spectral
shapes have been obtained in Ref.~\cite{ps1}, where the normalization
was chosen to match the observed differential flux at
$3\times10^{20}\,$eV. This normalization, however, 
leads to an overproduction of the
integral flux at higher energies, whereas above $10^{20}\,$eV,
the fits shown in Figs.~\ref{fig5} and~\ref{fig6} have
likelihood significances above 50\% (see Ref.~\cite{slsb} for
details) and are consistent with the integral flux above
$3\times10^{20}\,$eV estimated in Refs.~\cite{fe,agasa}.
The PP process on the CMB depletes the photon flux above 100 TeV, and the
same process on the IR/O background causes depletion of the photon flux in
the range 100 GeV--100 TeV, recycling the absorbed energies to
energies below 100 GeV through EM cascading (see Fig.~\ref{fig5}).
The predicted background is {\it not} very sensitive to
the specific IR/O background model, however~\cite{ahacoppi}.
The scenario in Fig.~\ref{fig5} obviously
obeys all current constraints within the normalization
ambiguities and is therefore quite viable. Note
that the diffuse $\gamma-$ray background measured by
EGRET~\cite{egret} up to 10 GeV puts a strong constraint on these
scenarios, especially if there is already a significant
contribution to this background from conventional
sources such as unresolved $\gamma-$ray blazars~\cite{muk-chiang}.
However, the $\gamma-$ray background constraint can be circumvented by
assuming that TDs or the decaying long lived X particles
do not have a uniform density throughout the
Universe but cluster within galaxies~\cite{bkv}. As can
also be seen, at
energies above 100 GeV, TD models are not significantly
constrained by observed $\gamma-$ray fluxes yet (see
Ref.~\cite{bs-rev} for more details on these measurements).

Fig.~\ref{fig6} shows results for the same TD scenario as in
Fig.~\ref{fig5}, but for a high EGMF $\sim 10^{-9}\,$G,
somewhat below the current upper limit~\cite{rkb}.
In this case, rapid synchrotron cooling of the initial cascade pairs quickly
transfers energy out of the UHE range. The UHE $\gamma-$ray flux then depends
mainly on the absorption length due to pair production and is typically
much lower~\cite{ABS,los}. (Note, though, that for $m_X\ga
10^{25}$ eV, the synchrotron radiation from these pairs
can be above $10^{20}\,$eV, and the UHE flux
is then not as low as one might expect.) We note, however,
that the constraints from the EGRET measurements do not
change significantly with the EGMF strength as long as the
nucleon flux is comparable
to the $\gamma-$ray flux at the highest energies, as is the
case in Figs.~\ref{fig5} and~\ref{fig6}.
The results of Ref.~\cite{slby} differ from
those of Ref.~\cite{ps1} which obtained more stringent
constraints on TD models because of the use of an older
fragmentation function from Ref.~\cite{hill}, and a stronger dependence
on the EGMF because of the use of a weaker EGMF which lead
to a dominance of $\gamma-$rays above $\simeq10^{20}\,$eV.

The energy loss and absorption lengths for UHE nucleons and photons
are short ($\la100$ Mpc). Thus, their predicted UHE fluxes are
independent of cosmological evolution. The $\gamma-$ray flux
below $\simeq10^{11}\,$eV, however, scales as the
total X particle energy release integrated over all redshifts
and increases with decreasing $p$~\cite{sjsb}. For
$m_X=2\times10^{16}\,$GeV,
scenarios with $p<1$ are therefore ruled
out (as can be inferred from Figs.~\ref{fig5} and ~\ref{fig6}), whereas
constant comoving injection models ($p=2$) are well within the
limits.

We now turn to signatures of TD models at UHE.
The full cascade calculations predict
$\gamma-$ray fluxes below 100 EeV that are a factor $\simeq3$
and $\simeq10$ higher than those obtained
using the CEL or absorption approximation often used in the
literature, in the case of strong and weak URB,
respectively. Again, this shows the importance
of non-leading particles in the development of unsaturated EM
cascades at energies below $\sim10^{22}\,$eV.
Our numerical simulations give a $\gamma$/CR flux ratio at
$10^{19}\,$eV of $\simeq0.1$. The experimental exposure
required to detect a $\gamma-$ray flux at that level is
$\simeq4\times10^{19}\,{\rm cm^2}\,{\rm sec}\,{\rm sr}$, about a
factor 10 smaller than the current total experimental exposure.
These exposures are well within reach of the
Pierre Auger Cosmic Ray Observatories~\cite{auger}, which may be able to
detect a neutral CR component down to a level of 1\% of the total
flux. In contrast, if the EGMF exceeds $\sim 10^{-11}\,$G, then UHE
cascading is inhibited, resulting in a lower UHE
$\gamma-$ray spectrum. In the $10^{-9}$ G scenario of Fig.~\ref{fig6},
the $\gamma$/CR flux ratio at $10^{19}\,$eV is $0.02$,
significantly lower than for no EGMF. 

It is clear from the above discussions that the predicted particle fluxes
in the TD scenario are currently uncertain to a large extent due to 
particle physics uncertainties (e.g., mass and decay modes of the X
particles, the quark fragmentation function, the nucleon fraction $f_N$,
and so on) as well as astrophysical uncertainties (e.g., strengths of the
radio and infrared backgrounds, extragalactic magnetic fields, etc.). 
More details on the dependence of the predicted UHE particle spectra and
composition on these particle physics and astrophysical
uncertainties are contained in Ref.~\cite{slby}. A detailed
study of the uncertainties involved in the propagation of
UHE nucleons, $\gamma-$rays, and neutrinos is currently
underway~\cite{kkss}.

We stress here that there are viable TD scenarios which
predict nucleon fluxes that are comparable to or even higher than
the $\gamma-$ray flux at all energies, even though $\gamma-$rays
dominate at production.
This occurs, e.g., in the case of high URB
and/or for a strong EGMF, and a nucleon fragmentation fraction of
$\simeq10\%$; see, for example, Fig.~\ref{fig6}. Some of these TD 
scenarios would therefore remain viable even if EHECR induced EAS
should be proven inconsistent with photon primaries (see,
e.g., Ref.~\cite{gamma}). This is in contrast to scenarios with
decaying massive dark matter in the Galactic halo which,
due to the lack of absorption, predict compositions directly
given by the fragmentation function, i.e. domination by
$\gamma-$rays.

The normalization procedure to the EHECR flux described above
imposes the constraint $Q^0_{\rm EHECR}\la10^{-22}\,{\rm eV}\,{\rm
cm}^{-3}\,{\rm sec}^{-1}$ within a factor of a
few~\cite{ps1,slby,slsc} for the total energy release rate $Q_0$
from TDs at the current epoch.
In most TD models, because of the unknown values of the
parameters involved, it is currently not
possible to calculate the exact value of $Q_0$ from first principles,
although it has been shown that the required values of $Q_0$ (in order to
explain the EHECR flux) mentioned above are quite possible for
certain kinds of TDs. Some cosmic
string simulations and the necklace scenario suggest that
defects may lose most of
their energy in the form of X particles and estimates of this
rate have been given~\cite{vincent,bv}. If that is the case, the
constraint on $Q^0_{\rm EHECR}$ translates via Eq.~(\ref{dotnx})
into a limit on the symmetry
breaking scale $\eta$ and hence on the mass $m_X$ of the X particle: 
$\eta\sim m_X\la10^{13}\,$GeV~\cite{wmgb}. Independently 
of whether or not this scenario explains EHECR, the EGRET measurement
of the diffuse GeV $\gamma-$ray background leads to a similar bound,
$Q^0_{\rm EM}\la2.2\times10^{-23}\,h
(3p-1)\,{\rm eV}\,{\rm cm}^{-3}\,{\rm sec}^{-1}$, which leaves
the bound on $\eta$ and $m_X$ practically unchanged.
Furthermore, constraints from limits on CMB distortions and light
element abundances from $^4$He-photodisintegration are
comparable to the bound from the directly observed
diffuse GeV $\gamma$-rays~\cite{sjsb}. That these crude
normalizations lead to values of $\eta$ in the right range
suggests that defect models require less fine tuning than
decay rates in scenarios of metastable massive dark matter.

\subsection{Results: Neutrino Fluxes}
As discussed in Sect.~4.1, in TD scenarios most of the energy is
released in the form of EM particles and neutrinos. If the X
particles decay into a quark and a lepton, the quark hadronizes
mostly into pions and the ratio of energy release into the
neutrino versus EM channel is $r\simeq0.3$.

Fig.~\ref{fig4} shows predictions of the total neutrino
flux for the same TD model on which
Fig.~\ref{fig5} is based. In the absence of neutrino
oscillations the electron neutrino and
anti-neutrino fluxes are about a factor of 2
smaller than the muon neutrino and anti-neutrino fluxes,
whereas the $\tau-$neutrino flux is in general negligible.
In contrast, if the
interpretation of the atmospheric neutrino deficit in terms
of nearly maximal mixing of muon and $\tau-$neutrinos proves
correct, the muon neutrino fluxes shown in Fig.~\ref{fig4} would
be maximally mixed with the $\tau-$neutrino fluxes. The TD flux
component clearly dominates above $\sim10^{19}\,$eV.

In order to translate neutrino fluxes into event rates,
one has to fold in the interaction cross sections with
matter. At UHEs these
cross sections are not directly accessible to laboratory
measurements. Resulting uncertainties therefore
translate directly to bounds on neutrino fluxes derived from,
for example, the non-detection of UHE muons produced in charged-current
interactions. In the following, we will assume the estimate
Eq.~(\ref{cccross2})
based on the Standard Model
for the charged-current muon-neutrino-nucleon
cross section $\sigma_{\nu N}$ if not indicated otherwise.

For an (energy dependent) ice or water equivalent acceptance
$A(E)$ (in units of volume times solid angle), one can obtain an
approximate expected rate of UHE muons produced by neutrinos
with energy $>E$, $R(E)$, by
multiplying $A(E)\sigma_{\nu N}(E)n_{\rm H_2O}$ (where 
$n_{\rm H_2O}$ is the nucleon density in water) with the
integral muon neutrino flux $\simeq Ej_{\nu_\mu}$. This can be used to
derive upper limits on diffuse neutrino fluxes from a
non-detection of muon induced events. Fig.~\ref{fig4} shows bounds
obtained from several experiments: The Frejus experiment
derived upper bounds for $E\ga10^{12}\,$eV from
their non-detection of almost horizontal muons with an energy
loss inside the detector of more than $140\,$MeV per radiation
length~\cite{frejus}. The AMANDA neutrino telescope has established
an upper limit in the TeV-PeV range~\cite{amanda}.
The Fly's Eye experiment derived
upper bounds for the energy range between $\sim10^{17}\,$eV and
$\sim10^{20}\,$eV~\cite{baltrusaitis} from the non-observation
of deeply penetrating particles. The NASA Goldstone radio telescope
has put an upper limit from the non-observation of
pulsed radio emission from cascades induced by neutrinos
above $\simeq10^{20}\,$eV in the lunar regolith.
The AKENO group has published an upper
bound on the rate of near-horizontal, muon-poor air
showers~\cite{nagano} (not shown in Fig.~\ref{fig4}). Horizontal
air showers created by electrons, muons or tau leptons that are in turn
produced by charged-current reactions of electron, muon or tau
neutrinos within the atmosphere have recently also been pointed out
as an important method to constrain or measure UHE neutrino
fluxes~\cite{auger-neut} with next generation detectors.

Clearly, the SLBY model is not only consistent with the constraints
discussed in Sect.~4.3, but also with all existing neutrino
flux limits within 2-3 orders of magnitude.
What, then, are the prospects of detecting UHE neutrino fluxes
predicted by TD models? In a $1\,{\rm km}^3\,2\pi\,$sr size
detector, the SLBY scenario from Fig.~\ref{fig4},
for example, predicts a muon-neutrino event rate
of $\simeq0.08\,{\rm yr}^{-1}$, and an electron neutrino event rate
of $\simeq0.05\,{\rm yr}^{-1}$ above $10^{19}\,$eV, where
``backgrounds'' from conventional sources should be negligible.
Further, the muon-neutrino event rate above 1 PeV should be
$\simeq0.6\,{\rm yr}^{-1}$, which could be interesting if
conventional sources produce neutrinos at a much smaller
flux level. Moreover, the neutrino flux around $10^{17}\,$eV
could have a slight enhancement due to neutrinos from muons
produced by interactions of UHE photons and electrons with
the CMB at high redshift~\cite{postma}, an effect that has
not been taken into account in the figures shown here.
Of course, above $\simeq100\,$TeV, instruments
using ice or water as detector medium, have to look at downward
going muon and electron events due to neutrino absorption in
the Earth. However, $\tau-$neutrinos obliterate this Earth
shadowing effect due to their regeneration from $\tau$
decays~\cite{halzen-saltzberg}. The presence of $\tau-$neutrinos,
for example, due to mixing with muon neutrinos, as suggested
by recent experimental results from Super-Kamiokande,
can therefore lead to an increased upward going event
rate~\cite{mannheim3}. As mentioned in Sect.~2, $\tau$ neutrinos
skimming the Earth at small angles below the horizon can also
lead to an increase of sensitivity of fluorescence
and ground array detectors~\cite{fargion,auger-tau,upgoing}.

For detectors based on the fluorescence technique such as the
HiRes~\cite{hires} and the Telescope Array~\cite{tel_array}
(see Sect.~2), the sensitivity to UHE neutrinos is often
expressed in terms of an effective aperture $a(E)$ which is
related to $A(E)$ by $a(E)=A(E)\sigma_{\nu N}(E)n_{\rm
H_2O}$. For the cross section of Eq.~(\ref{cccross2}), the
apertures given in Ref.~\cite{hires} for the HiRes correspond to
$A(E)\simeq3\,{\rm km}^3\times2\pi\,{\rm sr}$ for
$E\ga10^{19}\,$eV for muon neutrinos. The expected acceptance
of the ground array component of the Pierre Auger
project for horizontal
UHE neutrino induced events is $A(10^{19}\,{\rm eV})\simeq
20\,{\rm km}^3\,{\rm sr}$ and $A(10^{23}\,{\rm eV})\simeq
200\,{\rm km}^3\,{\rm sr}$~\cite{auger-neut}, with a duty cycle close to
100\%. We conclude that detection of neutrino
fluxes predicted by scenarios such as the SLBY scenario shown
in Fig.~\ref{fig4} requires running a detector of acceptance
$\ga10\,{\rm km}^3\times2\pi\,{\rm sr}$ over a period of a few
years. Apart from optical detection in air, water, or ice, other
methods such as acoustical and radio detection~\cite{ghs}
(see, e.g., the RICE project~\cite{rice} for the latter) or even
detection from space~\cite{owl,owl1,euso,airwatch} appear to be interesting
possibilities for detection concepts operating at such scales
(see Sect.~2). For example, the space based OWL/AirWatch satellite
concept would have an
aperture of $\simeq3\times10^6\,{\rm km}^2\,{\rm sr}$
in the atmosphere, corresponding to $A(E)\simeq6\times10^4
\,{\rm km}^3\,{\rm sr}$ for $E\ga10^{20}\,$eV,
with a duty cycle of $\simeq0.08$~\cite{owl,owl1}.
The backgrounds seem to be
in general negligible~\cite{ydjs,price}. As indicated by the
numbers above and by the projected sensitivities shown in
Fig.~\ref{fig4}, the Pierre Auger Project and especially the
space based AirWatch type projects should be capable of
detecting typical TD neutrino
fluxes. This applies to any detector of acceptance
$\ga100\,{\rm km}^3\,{\rm sr}$. Furthermore, a 100 day
search with a radio telescope of the NASA Goldstone type
for pulsed radio emission from cascades induced by neutrinos
or cosmic rays in the lunar regolith could reach
a sensitivity comparable or better to the Pierre Auger
sensitivity above $\sim10^{19}\,$eV~\cite{goldstone}.

A more model independent estimate~\cite{slsc} for the average
event rate $R(E)$ can be made if the underlying
scenario is consistent with observational nucleon and
$\gamma-$ray fluxes and the bulk of the energy is released above
the PP threshold on the CMB. Let us assume that the
ratio of energy injected into the neutrino versus EM channel is a
constant $r$. As discussed in Sect.~4.3, cascading effectively reprocesses
most of the injected EM energy into low
energy photons whose spectrum peaks at $\simeq10\,$GeV~\cite{ahacoppi}.
Since the ratio $r$ remains roughly unchanged during
propagation, the height of the
corresponding peak in the neutrino spectrum should
roughly be $r$ times the height of the low-energy
$\gamma-$ray peak, i.e., we have the condition
$\max_E\left[E^2j_{\nu_\mu}(E)\right]\simeq
r\max_E\left[E^2j_\gamma(E)\right].$ Imposing the observational
upper limit on the diffuse $\gamma-$ray flux around $10\,$GeV
shown in Fig.~\ref{fig4}, $\max_E\left[E^2j_{\nu_\mu}(E)\right]\la
2\times10^3 r \,{\rm eV}{\rm cm}^{-2}{\rm sec}^{-1}{\rm
sr}^{-1}$, then bounds the average diffuse neutrino rate above
PP threshold on the CMB, giving 
\begin{equation}
  R(E)\la0.34\,r\left[{A(E)\over1\,{\rm
  km}^3\times2\pi\,{\rm sr}}\right]
  \,\left({E\over10^{19}\,{\rm eV}}\right)^{-0.6}\,{\rm
  yr}^{-1}\quad(E\ga10^{15}\,{\rm eV})\,,\label{r2}
\end{equation}
assuming the Standard Model cross section Eq.~(\ref{cccross2}).
Comparing this with the flux bounds shown in Fig.~\ref{fig4}
results in an upper bound on $r$. For example, the Fly's Eye
bound translates into $r\la20(E/10^{19}\,{\rm eV})^{0.1}$.
We stress again that TD models are not subject to the Waxman
Bahcall bound because the nucleons produced are considerably
less abundant than and are not the primaries of produced
$\gamma-$rays and neutrinos.

\begin{figure}[htb]
\centerline{\hbox{\psfig{figure=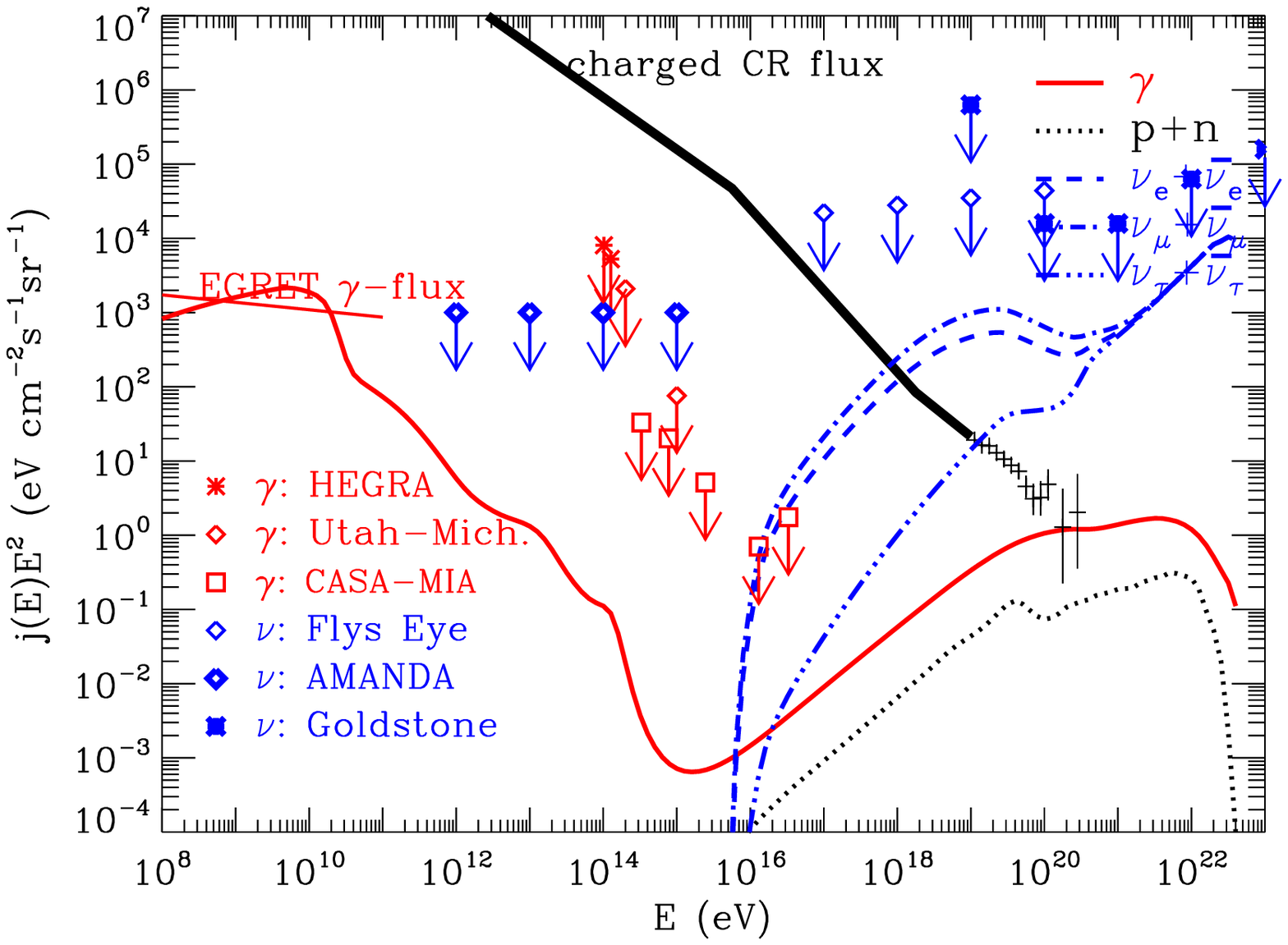,clip=true,width=0.9\textwidth}}}
\caption[...]{Flux predictions for a TD model characterized
by $p=1$, $m_X=10^{14}\,$GeV, with X particles exclusively
decaying into neutrino-antineutrino pairs of all flavors
(with equal branching ratio), assuming neutrino masses
$m_{\nu_e}=0.1\,$eV, $m_{\nu_\mu}=m_{\nu_\tau}=1\,$eV.
For neutrino clustering, an overdensity of $\simeq50$ over
a scale of $l_\nu\simeq5\,$Mpc was assumed. The calculation
assumed an intermediate URB estimate from Ref.~\cite{pb} and an
EGMF $\ll10^{-11}\,$G. The line key is as in Fig.~\ref{fig3}.
\label{fig7}}
\end{figure}

In typical TD models such as the one discussed above where
primary neutrinos are produced by pion decay,
$r\simeq0.3$. However, in TD scenarios with $r\gg1$ neutrino
fluxes are only limited by the condition that the {\it secondary}
$\gamma-$ray flux produced by neutrino interactions with the
relic neutrino background be below the experimental limits.
In this case the observed EHECR flux would be produced by
the Z-burst mechanism discussed in Sect.~3.1. An example for
such a scenario is given
by X particles exclusively decaying into neutrinos
(although this is not very likely in most particle physics
models, but see Ref.~\cite{slby} and Fig.~\ref{fig7} for a scenario
involving topological defects and Ref.~\cite{gk2} for a scenario
involving decaying superheavy relic particles, both of which explain the
observed EHECR events as secondaries of neutrinos interacting
with the primordial neutrino background). Such scenarios predict
appreciable event rates above $\sim10^{19}\,$eV in a km$^3$
scale detector, but require unrealistically strong clustering
of relic neutrinos (a homogeneous relic neutrino overdensity
would make the EGRET constraint only more severe
because neutrino interactions beyond $\sim50\,$Mpc
contribute to the GeV $\gamma-$ray background but not
to the UHECR flux). A detection would thus open the exciting
possibility to establish an experimental lower limit on $r$.
Being based solely on energy conservation,
Eq.~(\ref{r2}) holds regardless of whether or not the underlying
TD mechanism explains the observed EHECR events.

The transient neutrino event rate could be much higher than
Eq.~(\ref{r2}) in the direction to discrete sources which emit
particles in bursts. Corresponding pulses in the EHE
nucleon and $\gamma-$ray fluxes would only occur for sources
nearer than $\simeq100\,$Mpc and, in case of protons, would be
delayed and dispersed by deflection in Galactic and
extragalactic magnetic fields~\cite{wm,lsos}. The recent observation
of a possible clustering of CRs above $\simeq4\times10^{19}\,$eV by the
AGASA experiment~\cite{haya2} might suggest
sources which burst on a time scale $t_b\ll1\,$yr.
A burst fluence of $\simeq r\left[A(E)/1\,{\rm km}^3\times2\pi\,{\rm
sr}\right](E/10^{19}\,{\rm eV})^{-0.6}$ neutrino induced events
within a time $t_b$ could then be expected. Associated pulses
could also be observable in the ${\rm GeV}-{\rm TeV}$
$\gamma-$ray flux if the EGMF is smaller than
$\simeq10^{-15}\,$G in a significant fraction of extragalactic
space~\cite{wc}.

In contrast to roughly homogeneous sources and/or mechanisms
with branching ratios $r\gg1$, in scenarios involving
clustered sources such as metastable superheavy relic
particles decaying with $r\sim1$, the neutrino flux is
comparable to (not significantly larger than) the UHE photon
plus nucleon fluxes and thus comparable to the universal
cosmogenic flux marked ``$N\gamma$'' in Fig.~\ref{fig4}.
This can be understood because the neutrino flux is dominated
by the extragalactic contribution which scales with the
extragalactic nucleon and $\gamma-$ray contribution in exactly
the same way as in the unclustered case, whereas the extragalactic
contribution to the ``visible'' flux to be normalized to the
UHECR data is much smaller in the clustered case.
The resulting neutrino fluxes in these scenarios would
thus be much harder to detect even with next generation
experiments.

For recent compilations of UHE neutrino flux
predictions from astrophysical and TD sources see
Refs.~\cite{owl1,neutflux} and references therein.

\section{Conclusions}

Ultra-high energy cosmic rays have the potential to open a window
to and act as probes of new particle physics beyond the Standard Model
as well as processes occurring in the early Universe at energies close
to the Grand Unification scale. Even if their origin will turn out
to be attributable to astrophysical shock acceleration with no new
physics involved, they will still be witnesses of one of the most energetic
processes in the Universe. The future appears promising and
exciting due to the anticipated arrival of several large scale
experiments.

\section*{Acknowledgements}
I would like to thank all my collaborators in this research field
without whose efforts a good part of the work discussed here would not have
been possible.

\end{document}